\documentclass[twocolumn,longauthor,resetfootnote]{aastex7}

\newcommand{\teff}{$T_{\!\mbox{\scriptsize\it eff}}$}	
\newcommand{\logg}{log\,$g$}
\newcommand{\loggf}{log\,$g_{\!\mbox{\scriptsize\it F}}$}

\newcommand{\zsun}{$Z_\odot$}
\newcommand{\msun}{$M_\odot$}

\newcommand{\hii}{H\,{\sc ii}\rm}

\newcommand{\hiibf}{H\,{\footnotesize II}\bf}

\newcommand{\nii}{[N\,{\sc ii}]}

\newcommand{\oiii}{[O\,{\sc iii}]}

\newcommand{\oiiION}{O\,{\sc ii}}

\newcommand{\feii}{Fe\,{\sc ii}\rm}
\newcommand{\mgii}{Mg\,{\sc ii}\rm}
\newcommand{\tiii}{Ti\,{\sc ii}\rm}
\newcommand{\crii}{Cr\,{\sc ii}\rm}

\newcommand{\one}{\sc i\rm}

\newcommand{\cel}{{\sc cel}}
\newcommand{\rl}{{\sc rl}}
\newcommand{\eg}{{e.g.}}
\newcommand{\ie}{{i.e.}}

\newcommand{\trgb}{{\sc trgb}}
\newcommand{\bsg}{\mbox{\small BSG}}
\newcommand{\rsg}{\mbox{\small RSG}}

\newcommand{\hbeta}{H$\beta$}
\newcommand{\halpha}{H$\alpha$}
\newcommand{\lin}{$\,\lambda$}

\newcommand{\rtf}{$r_{25}$}

\newcommand{\oh}{12\,+\,log(O/H)}

\newcommand{\eosun}{$\epsilon_{\mathrm O,\odot}$}

\newcommand{\ohsun}{\mbox{12\,+\,log(O/H)$_\odot\,=\,$}}

\newcommand{\vs}{vs.}

\renewcommand{\eg}{\mbox{e.g.}}
\renewcommand{\ie}{\mbox{i.e.}}

\newcommand{\fwhm}{\mbox{\small FWHM}}

\newcommand{\ssc}{\mbox{\small SSC}}
\newcommand{\fglr}{\mbox{\sc fglr}}

\newcommand{\newhash}{%
	{\settoheight{\dimen0}{C}\kern-.05em \resizebox{!}{\dimen0}{\raisebox{\depth}{\#}}}}

\newcommand{\m}{\phantom{$-$}}			

\shorttitle{Blue supergiants in M101}
\shortauthors{Bresolin et al.}

\begin{document}

\title{Blue supergiants in the Pinwheel Galaxy M101: comparison with \hiibf\ region chemical abundances, spectroscopic distance and an independent determination of the Hubble constant}



\correspondingauthor{Fabio Bresolin}

\author[orcid=0000-0002-5068-9833]{Fabio Bresolin}
\affiliation{Institute for Astronomy, University of Hawaii, 2680 Woodlawn Drive, Honolulu, HI 96822, USA}
\email[]{bresolin@ifa.hawaii.edu}  

\author{Rolf-Peter Kudritzki}
\affiliation{Institute for Astronomy, University of Hawaii, 2680 Woodlawn Drive, Honolulu, HI 96822, USA}
\affil{Universit\"ats-Sternwarte, Fakult\"at f\"ur Physik, Ludwig-Maximilians Universit\"at M\"unchen, Scheinerstr. 1, D-81679 M\"unchen, Germany}
\email[]{kud@ifa.hawaii.edu}  

\author[orcid=0000-0002-9424-0501]{Miguel A. Urbaneja}
\affil{Universit\"at Innsbruck, Institut f\"ur Astro- und Teilchenphysik\\ 
	Technikerstr. 25/8, 6020 Innsbruck, Austria}
\email[]{Miguel.Urbaneja-Perez@uibk.ac.at}  

\author[orcid=0009-0001-5618-4326]{Eva Sextl}
\affil{Universit\"ats-Sternwarte, Fakult\"at f\"ur Physik, Ludwig-Maximilians Universit\"at M\"unchen, Scheinerstr. 1, D-81679 M\"unchen, Germany}
\email[]{sextl@usm.lmu.de}  

\author[orcid=0000-0002-6124-1196]{Adam G. Riess}
\affiliation{Department of Physics and Astronomy, Johns Hopkins University, Baltimore, MD 21218, USA}
\affiliation{Space Telescope Science Institute, 3700 San Martin Drive, Baltimore, MD 21218, USA}
\email[]{ariess@stsci.edu}  

\begin{abstract}
We present a quantitative spectroscopic study of 13 blue supergiant stars in the Pinwheel Galaxy M101, based on data obtained with the Low Resolution Imaging Spectrometer available at the Keck~I telescope. The average stellar metallicity decreases from $\sim$1.9\,\zsun\ near the center of the galaxy to $\sim$0.3\,\zsun\ at the optical outskirts. The galactocentric radial metallicity gradient is statistically consistent with previous studies of the gas-phase oxygen abundance  from \hii\ regions using the direct method. The \hii\ region-based Cepheid metallicities used by Riess et al.~in their determination of the Hubble constant $H_0$ are in substantial agreement with our measurements. The direct method gas-phase metallicities of the 18 star-forming galaxies we have analyzed so far, when adjusted upward for a mean $\sim$0.15 dex oxygen dust depletion factor, are in good agreement with those we infer from the supergiants, over a factor of 50 in metallicity. From the same data, we derive an expression for the metal-dependent depletion of oxygen in photoionized nebulae.
Utilizing the flux-weighted gravity--luminosity relationship (\fglr) of blue supergiants, we measure a distance to M101, $D=6.5\pm0.2$ Mpc ($\mu = 29.06 \pm 0.08$), which is within 1$\sigma$ from determinations based on the tip of the red giant branch and Cepheids. With M101 as a nearby SN~Ia host and using the observed standardized $B$-band magnitude of the supernova, our \fglr\ distance yields an independent value $H_0 = 72.5\pm4.6$ km\,s$^{-1}$\,Mpc$^{-1}$.
\end{abstract}



\section{Introduction}\label{sec:intro}
The Pinwheel Galaxy M101 (NGC~5457) is a grand design spiral galaxy that has been extensively studied due to its proximity ($D=6.7\pm0.1$~Mpc, \citealt{Riess:2024}), significant optical size in the sky (with an isophotal radius \rtf\,=\,8$'$, \citealt{Mihos:2013}), optimal viewing angle ($i=18^\circ$, \citealt{Walter:2008}), and numerous luminous star-forming regions (\citealt{Hodge:1990}). These characteristics have prompted several examinations of its gas-phase chemical abundances (\eg\ \citealt{Searle:1971, Kennicutt:1996, Kennicutt:2003, Bresolin:2007, Li:2013, Croxall:2016}) and distance estimates (refer to the recent compilation by \citealt{Huang:2024}).

Assessing the distance to M101 has been a pivotal topic in the early research of its most luminous blue stars. This line of inquiry began with \citet{Hubble:1926, Hubble:1936} and was further pursued by \citet{Sandage:1974}, who calculated a distance for this "nearest unobscured supergiant Sc" galaxy as $D=7.2\pm1$~Mpc. Despite the considerable uncertainty, this estimate aligns with the currently accepted value. Initial spectroscopic examinations of the blue supergiants of M101, which were aimed at understanding their inherent luminosities and the consequences for the galaxy's distance, include studies by \citet{Humphreys:1980} and \citet{Humphreys:1987}.

The distance to M101 has been successfully investigated by means of observations of Cepheid variables, red giant stars and other stellar standard candles with the Hubble Space Telescope (HST; \eg\ \citealt{Kelson:1996, Kennicutt:1998, Beaton:2019,Riess:2022}) and the James Webb Space Telescope (JWST;  \citealt{Riess:2024,Freedman:2025}). 
However, while studies of Wolf-Rayet stars (\citealt{Shara:2013, Pledger:2018}) and resolved massive stars as part of OB associations (\citealt{Bresolin:1996, Bresolin:1998}) are present in the M101 literature,
very little work has been carried out until recently on the characterization of individual massive blue supergiant stars in this galaxy. HST photometry of massive stars in M101 by \citet{Grammer:2013}
has led to a study of the star formation history across the disk of the galaxy (\citealt{Grammer:2014})
and to the spectroscopic followup of 31 luminous member stars (\citealt{Grammer:2015}), used primarily for stellar classification purposes.

Here we present new stellar spectroscopy in M101, as part of our study of the blue supergiants in nearby ($D < 8$~Mpc) star-forming galaxies, aimed at obtaining both their metallicities and the distance to their parent galaxies (see \citealt{Bresolin:2022, Urbaneja:2023, Kudritzki:2024} and references therein).
We show that a precise `stellar distance' for M101 is now feasible, a century after Hubble's first attempts to use luminous stars as extragalactic distance indicators in this and other targets. Moreover, the well-studied galactocentric gradient of the gas-phase metallicity of this galaxy, known since the pioneering work on abundance gradients in spiral galaxies by \citet{Searle:1971} and \citet{Smith:1975}, can be compared with the radial trend that we derive from young stars, affording a new test on the agreement between these two independent tracers of the metallicity of the youngest population observed in galaxies (see \citealt{Bresolin:2009a, Bresolin:2016}).

This work is organized as follows. The observational material, data reduction and spectral classification are presented in Section~2. The quantitative analysis, including the derivation of the stellar parameters and metallicities, follows in Section~3. The metallicity gradient, as determined from both blue supergiants and \hii\ regions, is discussed in Section~4. Section~5 looks at the results obtained for M101 in the context of similar studies we have carried out for additional nearby star-forming galaxies, and examines  the comparison between the stellar metallicities and the gas-phase abundances, as well as the galaxy mass-metallicity relation. We also derive an expression for the dependence of dust depletion of oxygen in \hii\ regions on metallicity, and compare the metallicities of the \bsg s  with those of the Cepheids used to measure the distance to M101.
The spectroscopic distance to M101 we derive from the stellar parameters is presented in Section~6, and
in Section~7 we combine it with the standardized magnitude of the Type Ia supernova 2011fe, in order to
calculate an independent value of the Hubble constant.
A summary of our results in Section~8 concludes our paper.

\section{Observations and data reduction}\label{sec:observations}
\subsection{Target selection}\label{subsec:selection}

Our blue supergiant candidates were drawn from the photometric catalog of massive stars in M101 by \citet[published online by \citealt{Grammer:2014a}]{Grammer:2013}, based on \textit{Hubble Space Telescope} (HST) Advanced Camera for Survey images obtained in programs GO-9490 (PI: Kuntz) and GO-9492 (PI: Bresolin) with the F435W, F555W and F814W filters. The stellar photometry, measured with {\sc Dolphot} (\citealt{Dolphin:2000, Dolphin:2016}), was converted to Johnson-Cousins \textit {BVI}  magnitudes with the transformations provided by \citet{Sirianni:2005}.

We selected our targets by imposing magnitude ($18.8<V<21.5$) and color index  ($-0.1 < B-V < 0.35$, $-0.1 < V-I < 0.3$) restrictions, in order to isolate potential blue supergiants. The final list of targets was obtained by rejecting stars located in the most crowded regions and those with bright neighbors, as determined from visual inspection of the HST images.

M101 has a relatively large optical extension on the sky: its asymmetric spiral structure can be traced out to approximately 15 arcmin from the center.
The HST data, and therefore our blue supergiant study, cover only the central portion of the galaxy, out to about one isophotal radius. We adopt an isophotal radius \rtf\,=\,8.0 arcmin from \citet{Mihos:2013}, rather than the much larger RC3 (\citealt{de-Vaucouleurs:1991}) value of 14.4 arcmin that has been commonly used in investigations of the galactocentric abundance gradient of M101 from \hii\ regions (\eg~\citealt{Kennicutt:2003, Croxall:2016}). When necessary, in our analysis we have rescaled the galactocentric distances adopted by previous authors.

\begin{figure*}
	\center
	\includegraphics[width=1\textwidth]{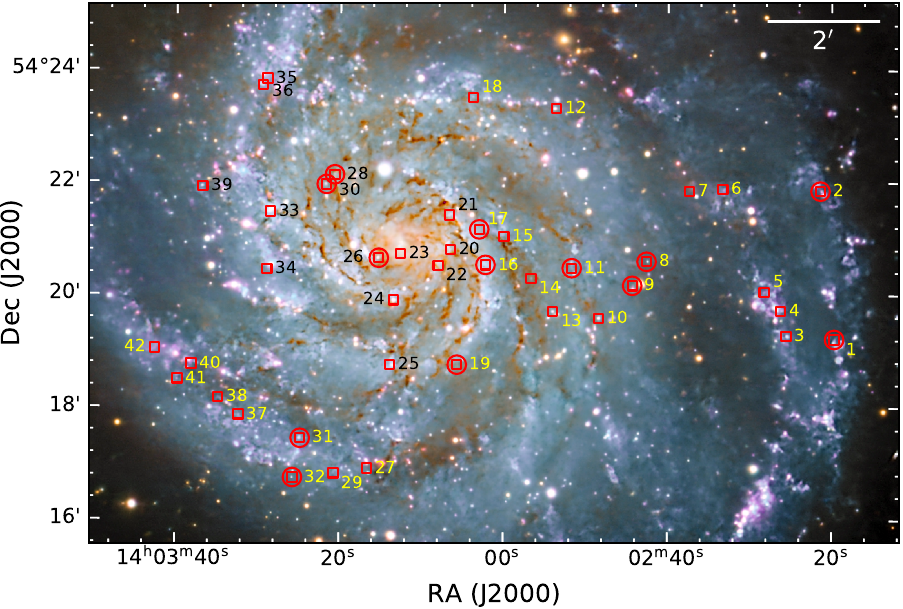}
	\caption{Positions of the spectroscopic targets. The stars analyzed in Section~\ref{sec:quantitative} are identified by the double square+circle markers. Different label colors are used for visibility purposes. Color image of M101 courtesy of Paolo Speggiorin (Specola Don Paolo Chiavacci -- Crespano del Grappa, Italy).
 }\label{fig:targets}\bigskip
\end{figure*}

\subsection{Spectroscopy}

The spectra presented in this work were acquired with the blue channel of the Low Resolution Imaging Spectrometer (LRIS, \citealt{Oke:1995}) at the Keck~I telescope, on two separate observing runs, 2023 May 13 (run 1) and 2024 April 6-7 (run 2), with two distinct slit mask setups. The data, obtained with the 600/4000 grism and a 1\farcs2 slit width, cover the approximate wavelength range 3300--5600~\AA\ with a 
\fwhm\ spectral resolution of about 5\,\AA. Individual exposures of 2700\,s were secured, for total integration times of 6.8\,h (run 1) and 8.2\,h (run 2). The weather conditions were excellent during run 1, with clear sky and seeing of 0.7 arcsec, while the two nights of run 2 were affected by intermittent clouds and variable atmospheric transparency, with seeing around 1.1 arcsec.

Both our masks comprised 21 slits (one slit per stellar target), with no object in common between the two setups. Although more \bsg\ candidates were available, the slit mask design had to account for tradeoffs involving several factors: sufficient spectral coverage for the analysis, spatial separation between targets and avoidance of spectral overlaps, availability of background regions, wide radial coverage, and overall telescope time request. The positions of the targets are identified in Figure~\ref{fig:targets}, and we report in Table~\ref{table:1} (sorted by increasing RA) their celestial coordinates, galactocentric distances (both in \rtf\ units and in kpc), $V$ magnitudes, $B-V$ and $V-I$ color indices, and spectral types (Sec.~\ref{sec:classification}). It can be seen that the spectroscopic targets lie in the magnitude range $19.9 \leq V \leq 21.5$.

The data reduction was carried out with {\sc iraf/p}y{\sc raf} and included the usual steps of bias subtraction, flat field correction and wavelength calibration. The spectral extractions for individual exposures were registered with the aid of sky lines and later combined. Finally, we rectified the combined spectra using low-order polynomials, and applied a radial velocity correction based on stellar lines in order to work in the rest frame.
After completion of these steps, the average signal-to-noise ratio of the coadded spectra ranges between 40 and 220 (median: 140). 

\begin{deluxetable}{cCCcccrrc}
	\tabletypesize{\footnotesize}	
	\tablecaption{Spectroscopic targets  \label{table:1}}
	
	\tablehead{
		\colhead{ID}	     		&
		\colhead{R.A.}	 			&
		\colhead{Decl.}	 			&
		\colhead{$r/r_{25}$}	 	&
		\colhead{$r$}               &
		\colhead{$V$}				&
		\colhead{$B-V$}				&
		\colhead{$V-I$}				&
		\colhead{Spectral}\\[-2.5ex]          
		\colhead{}       			&
		\colhead{(J2000.0)}       	&
		\colhead{(J2000.0)}       	&
		\colhead{}                  &
		\colhead{(kpc)}             &
		\colhead{(mag)}            	& 	
		\colhead{(mag)}            	& 
		\colhead{(mag)}             & 															
		\colhead{type}\\[-5ex] } 
	\colnumbers
	\startdata
	\\[-4.5ex]
{\bf 01} &  {\bf 14\; 02\; 19.69}  &  {\bf 54\; 19\; 13.6}  & {\bf    1.01}  & {\bf    15.6}  &  {\bf 21.00}  &  $\bf 0.19$  &  $\bf 0.13$   &   {\bf A2               } \\[-0.4ex]  
{\bf 02} &  {\bf 14\; 02\; 21.39}  &  {\bf 54\; 21\; 51.7}  & {\bf    0.97}  & {\bf    15.1}  &  {\bf 20.98}  &  $\bf 0.22$  &  $\bf 0.28$   &   {\bf A5               } \\[-0.4ex]  
03        &  14\; 02\; 25.55  &  54\; 19\; 17.2  &    0.90  &    13.9  &  20.85  &  $ 0.15$  &  $ 0.14$   &   A1                \\[-0.4ex]  
04        &  14\; 02\; 26.24  &  54\; 19\; 44.0  &    0.88  &    13.6  &  20.69  &  $ 0.09$  &  $-0.07$   &   B5                \\[-0.4ex]  
05        &  14\; 02\; 28.23  &  54\; 20\; 04.0  &    0.83  &    12.9  &  21.44  &  $ 0.14$  &  $ 0.14$   &   (A2,\,\hii)       \\[-0.4ex]  
06        &  14\; 02\; 33.31  &  54\; 21\; 53.6  &    0.75  &    11.7  &  21.35  &  $ 0.08$  &  $ 0.01$   &   B5/B6             \\[-0.4ex]  
07        &  14\; 02\; 37.36  &  54\; 21\; 51.7  &    0.68  &    10.5  &  21.34  &  $ 0.01$  &  $-0.23$   &   (early B,\,\hii)  \\[-0.4ex]  
{\bf 08} &  {\bf 14\; 02\; 42.59}  &  {\bf 54\; 20\; 36.4}  & {\bf    0.56}  & {\bf     8.7}  &  {\bf 20.63}  &  $\bf 0.15$  &  $\bf 0.16$   &   {\bf A2               } \\[-0.4ex]  
{\bf 09} &  {\bf 14\; 02\; 44.33}  &  {\bf 54\; 20\; 10.6}  & {\bf    0.53}  & {\bf     8.3}  &  {\bf 21.09}  &  $\bf 0.20$  &  $\bf 0.25$   &   {\bf A2               } \\[-0.4ex]  
10        &  14\; 02\; 48.46  &  54\; 19\; 35.8  &    0.48  &     7.4  &  20.11  &  $ 0.19$  &  $ 0.22$   &   A2,\,\hii         \\[-0.4ex]  
{\bf 11} &  {\bf 14\; 02\; 51.72}  &  {\bf 54\; 20\; 29.1}  & {\bf    0.39}  & {\bf     6.1}  &  {\bf 21.16}  &  $\bf 0.20$  &  $\bf 0.17$   &   {\bf A2               } \\[-0.4ex]  
12        &  14\; 02\; 53.64  &  54\; 23\; 19.2  &    0.48  &     7.4  &  21.42  &  $ 0.23$  &  $ 0.26$   &   A2/A3,\,\hii      \\[-0.4ex]  
13        &  14\; 02\; 54.07  &  54\; 19\; 42.9  &    0.37  &     5.8  &  20.73  &  $ 0.04$  &  $ 0.06$   &   B7/B8             \\[-0.4ex]  
14        &  14\; 02\; 56.68  &  54\; 20\; 18.2  &    0.30  &     4.7  &  21.16  &  $ 0.09$  &  $ 0.12$   &   composite         \\[-0.4ex]  
15        &  14\; 03\; 00.02  &  54\; 21\; 02.9  &    0.24  &     3.7  &  21.28  &  $ 0.16$  &  $ 0.26$   &   (BA\,+\,WN,\,\hii) \\[-0.4ex]  
{\bf 16} &  {\bf 14\; 03\; 02.24}  &  {\bf 54\; 20\; 32.3}  & {\bf    0.20}  & {\bf     3.1}  &  {\bf 21.17}  &  $\bf 0.10$  &  $\bf 0.14$   &   {\bf A2               } \\[-0.4ex]  
{\bf 17} &  {\bf 14\; 03\; 03.02}  &  {\bf 54\; 21\; 10.1}  & {\bf    0.18}  & {\bf     2.8}  &  {\bf 21.42}  &  $\bf 0.13$  &  $\bf 0.13$   &   {\bf A2               } \\[-0.4ex]  
18        &  14\; 03\; 03.83  &  54\; 23\; 30.9  &    0.37  &     5.8  &  20.58  &  $ 0.25$  &  $ 0.23$   &   A7 (\hii)         \\[-0.4ex]  
{\bf 19} &  {\bf 14\; 03\; 05.68}  &  {\bf 54\; 18\; 45.7}  & {\bf    0.30}  & {\bf     4.7}  &  {\bf 21.37}  &  $\bf 0.05$  &  $\bf 0.08$   &   {\bf B8               } \\[-0.4ex]  
20        &  14\; 03\; 06.55  &  54\; 20\; 48.2  &    0.11  &     1.8  &  20.84  &  $ 0.15$  &  $ 0.21$   &   (early B\,+\,H emission) \\[-0.4ex]  
21        &  14\; 03\; 06.66  &  54\; 21\; 25.2  &    0.13  &     2.0  &  21.11  &  $ 0.27$  &  $ 0.29$   &   composite         \\[-0.4ex]  
22        &  14\; 03\; 08.01  &  54\; 20\; 31.1  &    0.10  &     1.5  &  21.14  &  $ 0.17$  &  $ 0.23$   &   (A2/composite)    \\[-0.4ex]  
23        &  14\; 03\; 12.61  &  54\; 20\; 43.7  &    0.03  &     0.4  &  20.40  &  $ 0.13$  &  $ 0.20$   &   composite         \\[-0.4ex]  
24        &  14\; 03\; 13.44  &  54\; 19\; 54.4  &    0.13  &     2.1  &  19.90  &  $-0.02$  &  $ 0.06$   &   early A/composite \\[-0.4ex]  
25        &  14\; 03\; 13.86  &  54\; 18\; 45.5  &    0.28  &     4.3  &  21.34  &  $ 0.15$  &  $ 0.25$   &   (A2)              \\[-0.4ex]  
{\bf 26} &  {\bf 14\; 03\; 15.26}  &  {\bf 54\; 20\; 39.3}  & {\bf    0.06}  & {\bf     1.0}  &  {\bf 20.70}  &  $\bf 0.28$  &  $\bf 0.28$   &   {\bf A7               } \\[-0.4ex]  
27        &  14\; 03\; 16.61  &  54\; 16\; 54.8  &    0.52  &     8.1  &  21.05  &  $ 0.20$  &  $ 0.21$   &   (early B,\,\hii)  \\[-0.4ex]  
{\bf 28} &  {\bf 14\; 03\; 20.66}  &  {\bf 54\; 22\; 08.0}  & {\bf    0.21}  & {\bf     3.3}  &  {\bf 21.21}  &  $\bf 0.18$  &  $\bf 0.21$   &   {\bf B8/B9            } \\[-0.4ex]  
29        &  14\; 03\; 20.68  &  54\; 16\; 49.6  &    0.55  &     8.6  &  21.48  &  $ 0.05$  &  $ 0.08$   &   (early B,\,\hii)  \\[-0.4ex]  
{\bf 30} &  {\bf 14\; 03\; 21.66}  &  {\bf 54\; 21\; 57.7}  & {\bf    0.21}  & {\bf     3.3}  &  {\bf 21.07}  &  $\bf 0.09$  &  $\bf 0.14$   &   {\bf A3               } \\[-0.4ex]  
{\bf 31} &  {\bf 14\; 03\; 24.77}  &  {\bf 54\; 17\; 26.7}  & {\bf    0.51}  & {\bf     7.9}  &  {\bf 20.90}  &  $\bf 0.15$  &  $\bf 0.21$   &   {\bf A2               } \\[-0.4ex]  
{\bf 32} &  {\bf 14\; 03\; 25.68}  &  {\bf 54\; 16\; 44.9}  & {\bf    0.60}  & {\bf     9.3}  &  {\bf 20.66}  &  $\bf 0.22$  &  $\bf 0.27$   &   {\bf A3               } \\[-0.4ex]  
33        &  14\; 03\; 28.50  &  54\; 21\; 28.4  &    0.30  &     4.7  &  20.94  &  $ 0.27$  &  $ 0.24$   &   (composite,\,\hii) \\[-0.4ex]  
34        &  14\; 03\; 28.91  &  54\; 20\; 26.9  &    0.32  &     4.9  &  21.49  &  $ 0.10$  &  $ 0.15$   &   (early A,\,\hii)  \\[-0.4ex]  
35        &  14\; 03\; 28.91  &  54\; 23\; 50.2  &    0.47  &     7.3  &  21.24  &  $ 0.06$  &  $ 0.06$   &   composite         \\[-0.4ex]  
36        &  14\; 03\; 29.44  &  54\; 23\; 43.4  &    0.46  &     7.2  &  21.05  &  $ 0.13$  &  $ 0.13$   &   (early A,\,\hii)  \\[-0.4ex]  
37        &  14\; 03\; 32.31  &  54\; 17\; 51.8  &    0.55  &     8.6  &  21.38  &  $ 0.12$  &  $ 0.20$   &   composite         \\[-0.4ex]  
38        &  14\; 03\; 34.78  &  54\; 18\; 10.2  &    0.56  &     8.7  &  21.47  &  $ 0.05$  &  $ 0.02$   &   B2                \\[-0.4ex]  
39        &  14\; 03\; 36.78  &  54\; 21\; 55.4  &    0.47  &     7.2  &  20.43  &  $ 0.06$  &  $ 0.01$   &   B4 (\hii)         \\[-0.4ex]  
40        &  14\; 03\; 38.08  &  54\; 18\; 45.6  &    0.57  &     8.8  &  20.96  &  $ 0.21$  &  $ 0.26$   &   A0/A2 (\hii)      \\[-0.4ex]  
41        &  14\; 03\; 39.78  &  54\; 18\; 29.7  &    0.61  &     9.5  &  21.36  &  $ 0.04$  &  $-0.03$   &   composite         \\[-0.4ex]  
42        &  14\; 03\; 42.56  &  54\; 19\; 02.6  &    0.62  &     9.7  &  21.03  &  $ 0.09$  &  $ 0.07$   &   (early B,\,\hii)  \\[-0.4ex]  
	\\[-2.5ex]
	\enddata
	\tablecomments{Stars identified with boldface characters are those analyzed in Section~\ref{sec:quantitative}. The photometric measurements were taken from \citet{Grammer:2013}. The galactocentric radii in columns (4) and (5) were calculated adopting the following parameters: $i=18^\circ$, $\mathrm{PA}=39^\circ$ (\citealt{Walter:2008}), $r_{25}=8\farcm0$ (\citealt{Mihos:2013}), $D=6.67$~Mpc (\citealt{Riess:2024}). The coordinates of the galactic center were taken from the NASA/IPAC Extragalactic Database (NED): RA(J2000) = 14:03:12.544, DEC(J2000) = 54:20:56.22. We use the ``\hii" label in column (9) when the presence of nebular emission lines prevents any quantitative analysis or affects the spectral classification. Uncertain spectral types are indicated by the use of parentheses.}
\end{deluxetable}


\begin{figure*}
	\center \includegraphics[width=1\textwidth]{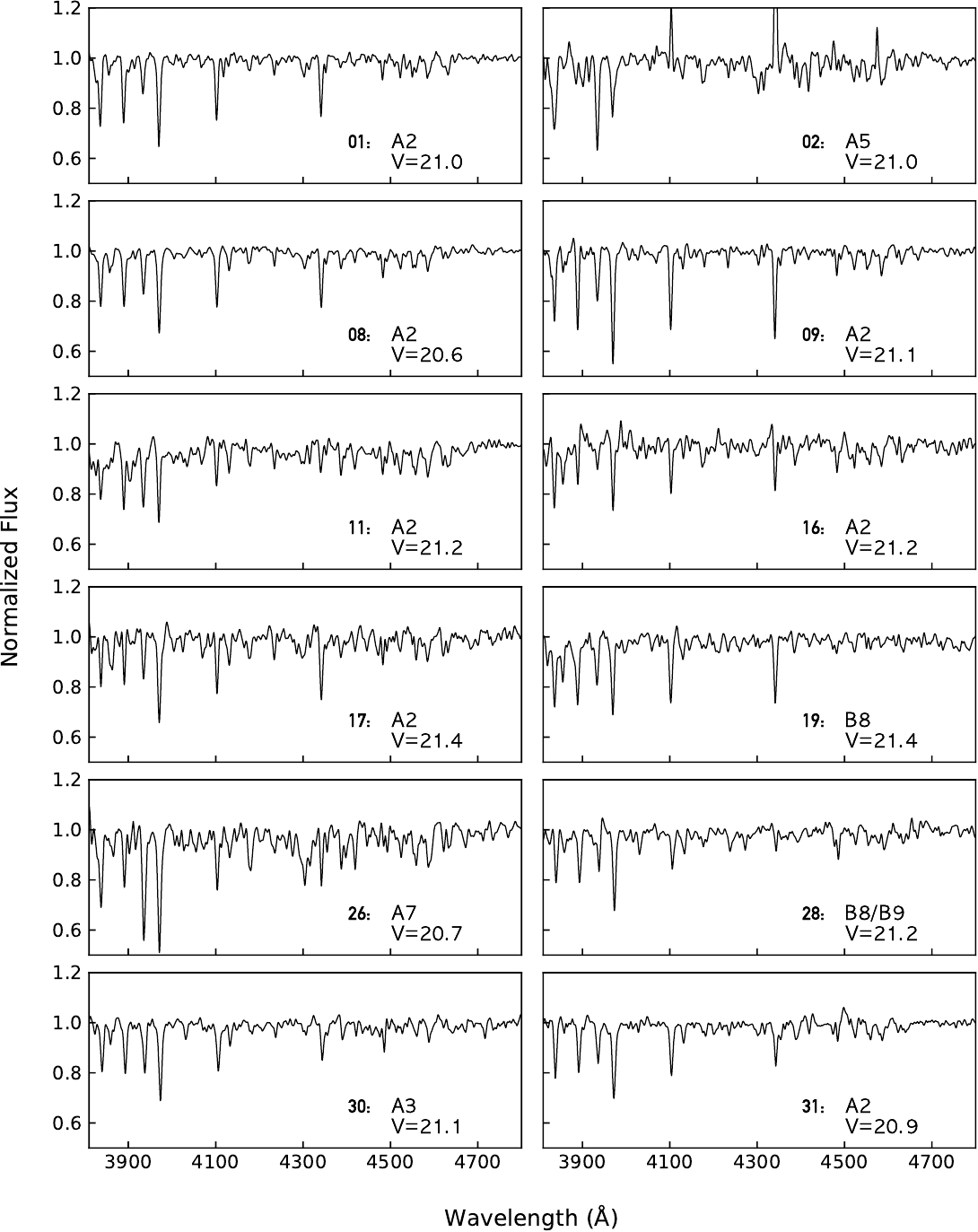}\medskip
	\caption{Spectra of 12 of the 13 targets analyzed. In each panel we report ID, spectral type and $V$ magnitude from Table~\ref{table:1}.}\label{fig:spectra}
\end{figure*}

\subsection{Spectral classification}\label{sec:classification}

Although assigning a spectral classification to our targets is not necessary for the quantitative analysis performed in Section~\ref{sec:quantitative}, we still estimated the spectral types of the stars we observed by following the MK criteria and the monograph by \citet{Gray:2009}, as well as visually comparing our rectified spectra with those of Galactic stars of known spectral types. Our results are included in Column (9) of Table~\ref{table:1}. 
At the distance of M101, especially in the vicinity of star-forming regions, the contamination by nebular emission and stellar neighbors is often unavoidable, despite our efforts to minimize their effects. This problem was already well recognized by \citet{Humphreys:1987a} in their study of the brightest blue supergiant candidates in NGC~2403, M81 and M101.
As a consequence, a large fraction of the targets in Table~\ref{table:1} are marked as being significantly affected by emission lines from ionized gas (the ``\hii" label is used)
or as ``composite". In several cases we include the assigned spectral types within parentheses, in order to indicate that they are quite uncertain. This appears to be due to a combination of faint magnitudes, increasing number of  ill-defined spectral lines for the cooler stars, and the potential presence of contaminating stellar neighbors.

It can be seen from Table~\ref{table:1} that our photometric selection successfully isolated the intended blue supergiants, with spectral types ranging mainly from mid B to late A. Our quantitative analysis focuses on mostly \hii-uncontaminated stars in the spectral range B8-A7, for which our models are optimized, the brightest being found around $V\simeq20.6$. Our brightest target, $\#24$ ($V=19.9$), unfortunately displays a composite spectrum. On the other hand, the stars we were able to analyze cover rather uniformly a wide range of galactocentric distances: $r/$\rtf\,=\,0.06-1.01 ($r$\,=\,1.0-15.6~kpc). This allows us to measure a reliable metallicity gradient (Section~\ref{sec:gradient}).

We have two objects in common with the study by \citet{Grammer:2015}, who obtained Multi Mirror Telescope spectra of bright stars in M101: their targets 
9490\underline{\hspace{0.5em}}c2\underline{\hspace{0.5em}}120 
(classified as a late A supergiant) and 9490\underline{\hspace{0.5em}}02\underline{\hspace{0.5em}}1086
(late B). 
We rejected both stars from our analysis: although we classify the first object (star 10 in Table~\ref{table:1}) as an early A supergiant, the contamination of the Balmer lines by nebular emission is so severe, even at high order, that it prevents us from determining the stellar gravity from model fits (Section~\ref{sec:quantitative}). The spectrum of the second object (our star 24) displays features characteristic of early-A stars, but it is likely a composite. Its spatial profile is also somewhat broader than that of a single star.

\section{Quantitative analysis}\label{sec:quantitative}

\begin{figure*}
	\center
	\includegraphics[width=0.8\textwidth]{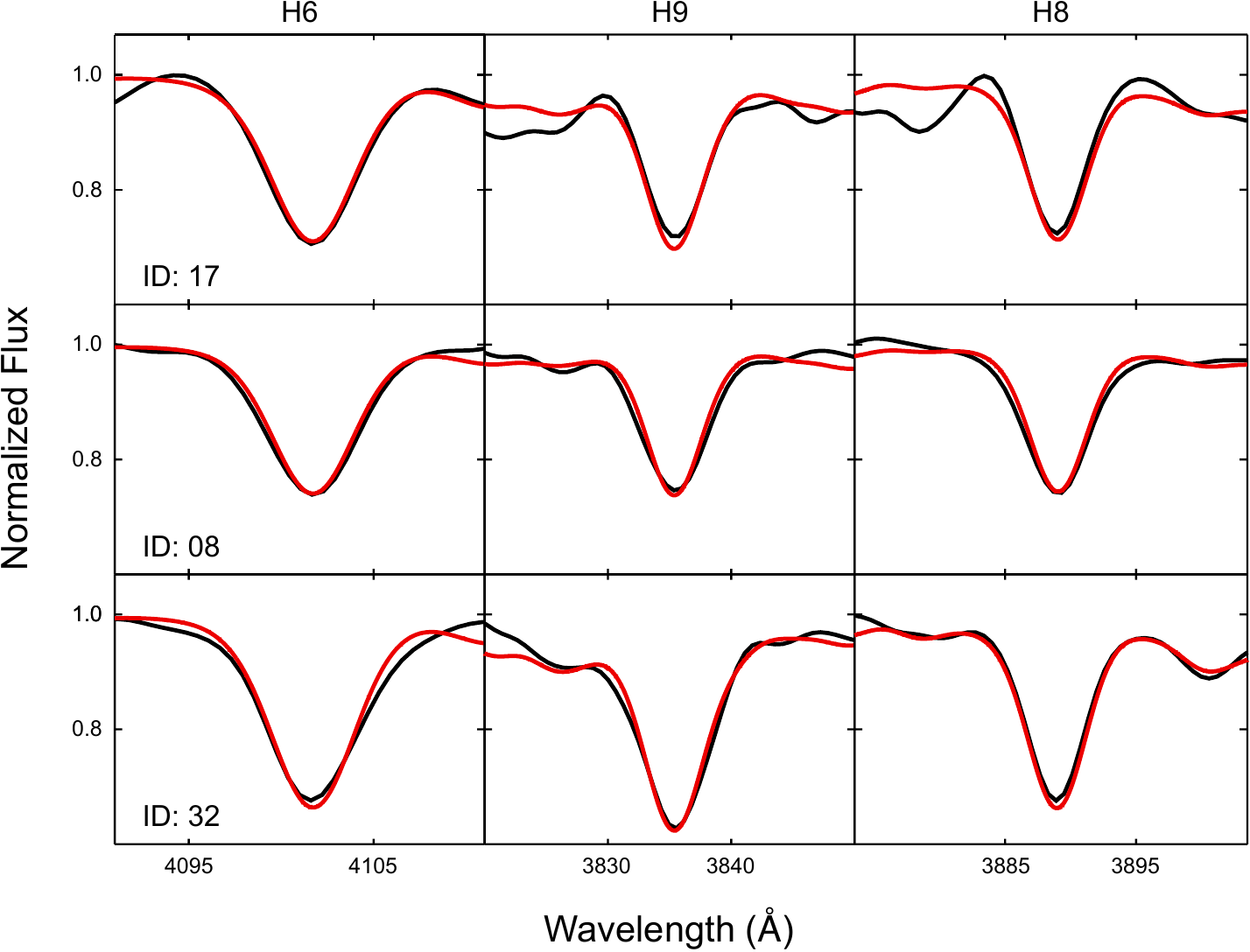}
	\caption{Fits of the adopted model spectra (red lines) to the observed H6, H9 and H8 Balmer line profiles (black lines) for targets ID 17 (top), 08 (middle) and 32 (bottom). 
	}\label{fig:Balmer-fits}\bigskip
\end{figure*}

\begin{deluxetable*}{cccccccc}
	\tabletypesize{\footnotesize}	
	\tablecaption{Stellar parameters\label{table:2}}
	
	\tablehead{
		\colhead{\phn ID \phn}	     		&
		\colhead{S/N}               &
		\colhead{\teff}	 			&
		\colhead{\logg}	 			&
		\colhead{$[Z]$}			    &
		\colhead{\phn\phn\phn \loggf \phn\phn\phn}			&
		\colhead{m$_{bol}$}         &
		\colhead{$E(B-V)$}\\[-2.ex]
		\colhead{}       			&
		\colhead{}       			&
		\colhead{(K)}       	&
		\colhead{(cgs)}       	&
		\colhead{(dex)}					& 	
		\colhead{(cgs)}					& 
		\colhead{(mag)}				& 														
		\colhead{(mag)}\\[-5ex] }
	
	\colnumbers
	\startdata
	\\[-4.5ex]
	01 & 47 &  $8750\pm50$  & $1.10\pm 0.05$ &  $-0.40\pm 0.03$ 			&  $1.332\;^{0.062}_{0.059}$ & $20.611\pm 0.034$ &   $0.088\pm 0.008$ \\[-0.ex]		
	02 & 51 &  $7900\pm50$  & $0.65\pm 0.05$ &  $-0.53\pm 0.06$ 			&  $1.059\;^{0.087}_{0.091}$ & $20.646\pm 0.090$ &   $0.121\pm 0.024$ \\[-0.ex]		
	08 & 77 &  $9125\pm175$ & $1.11\pm 0.05$ &  $-0.16\pm 0.07$ 			&  $1.270\;^{0.073}_{0.069}$ & $20.187\pm 0.060$ &   $0.081\pm 0.010$ \\[-0.ex]		
	09 & 45 &  $9375\pm200$ & $1.51\pm 0.05$ &  $-0.12\pm 0.10$ 			&  $1.620\;^{0.081}_{0.086}$ & $20.311\pm 0.049$ &   $0.192\pm 0.009$ \\[-0.ex]		
	11 & 41 &  $9200\pm150$ & $1.18\pm 0.05$ &  $\phantom{-}0.29\pm 0.06$ 	&  $1.327\;^{0.076}_{0.074}$ & $20.629\pm 0.053$ &   $0.125\pm 0.010$ \\[-0.ex]		
	16 & 31 &  $9400\pm250$ & $1.17\pm 0.05$ &  $-0.10\pm 0.10$ 			&  $1.279\;^{0.073}_{0.075}$ & $20.728\pm 0.072$ &   $0.068\pm 0.010$ \\[-0.ex]		
	17 & 33 &  $9200\pm200$ & $1.18\pm 0.05$ &  $\phantom{-}0.15\pm 0.10$ 	&  $1.328\;^{0.081}_{0.079}$ & $21.036\pm 0.065$ &   $0.073\pm 0.012$ \\[-0.ex]		
	19 & 30 & $10200\pm300$ & $1.50\pm 0.05$ &  $\phantom{-}0.10\pm 0.12$ 	&  $1.461\;^{0.065}_{0.069}$ & $20.834\pm 0.071$ &   $0.070\pm 0.011$ \\[-0.ex]		
	26 & 39 &  $8200\pm100$ & $0.84\pm 0.05$ &  $\phantom{-}0.21\pm 0.07$ 	&  $1.185\;^{0.096}_{0.098}$ & $20.269\pm 0.090$ &   $0.141\pm 0.022$ \\[-0.ex]		
	28 & 39 & $10950\pm250$ & $1.49\pm 0.05$ &  $\phantom{-}0.28\pm 0.07$ 	&  $1.333\;^{0.057}_{0.057}$ & $20.112\pm 0.062$ &   $0.200\pm 0.009$ \\[-0.ex]		
	30 & 43 &  $9700\pm200$ & $1.29\pm 0.05$ &  $\phantom{-}0.07\pm 0.12$ 	&  $1.340\;^{0.066}_{0.067}$ & $20.560\pm 0.057$ &   $0.078\pm 0.009$ \\[-0.ex]		
	31 & 46 &  $9150\pm200$ & $1.17\pm 0.05$ &  $-0.13\pm 0.08$ 			&  $1.322\;^{0.079}_{0.076}$ & $20.363\pm 0.064$ &   $0.113\pm 0.011$ \\[-0.ex]		
	32 & 57 &  $8225\pm75$  & $0.86\pm 0.05$ &  $-0.27\pm 0.03$ 			&  $1.194\;^{0.081}_{0.081}$ & $20.257\pm 0.066$ &   $0.117\pm 0.015$ \\[-0.ex]		
	\\[-2.5ex]
	\enddata
	\tablecomments{The \logg\ uncertainty is for a fixed \teff\ value. }
\end{deluxetable*}


\begin{figure}
	\center
	\includegraphics[width=1\columnwidth]{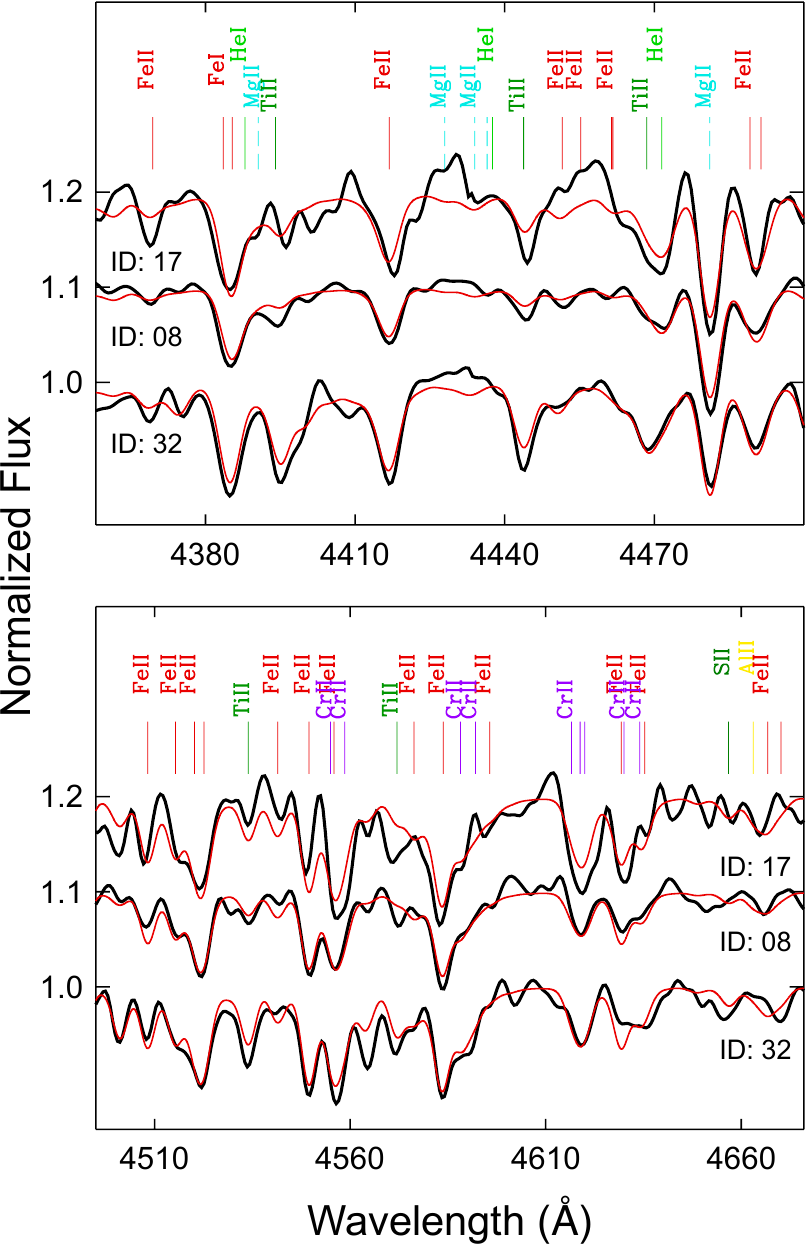}
	\caption{Fits of the adopted model spectra (in red) to the observed metal lines (in black) for targets ID 17, 08 and 32, shifted vertically by 0.1 normalized units from each other, in two distinct wavelength windows. We mark the positions of the main lines included in the synthetic spectra calculations.
	}\label{fig:metal-fits}\bigskip
\end{figure}

\subsection{Main stellar parameters}
For 13 of the stars presented in Section~\ref{sec:observations} (identified with boldface characters in Table~\ref{table:1}) we are able to obtain the fundamental stellar parameters: effective temperature (\teff), surface gravity (\logg) and metallicity ($[Z]$). The latter is the logarithmic mass fraction of metals, relative to the solar value \zsun\,=\,0.014 (\citealt{Asplund:2009}). The spectra of 12 of these targets are illustrated in Figure~\ref{fig:spectra}, while Table~\ref{table:2} summarizes the stellar parameters that we derived, and includes the signal-to-noise ratio (S/N) of the spectra.

We follow the same procedure that we adopted in previous works of our series (see \citealt{Urbaneja:2023} and \citealt{Kudritzki:2024} for recent examples), based on a comparison between the observed, rectified spectra of our targets with synthetic spectra built from a dense grid of non-LTE model atmospheres (\citealt{Przybilla:2006, Kudritzki:2008, Kudritzki:2012}) comprising effective temperatures from 7900\,K to 15000\,K, with gravities ranging between 0.8 and 3.0 dex (in cgs units, the exact lower and upper limits depending on \teff; see Figure~1 in \citealt{Kudritzki:2008} for details on the grid points). The metallicity $[Z]$ ranges between $-1.45$ and +0.5 dex. The synthetic spectra are convolved to the LRIS resolution. We note that at the low resolution of LRIS we are not able to determine rotational velocities, which for our targets are expected to be on the order of 50~km\,s$^{-1}$. As detailed comparisons with high-resolution spectra have shown \citep{Urbaneja:2017}, the effects of low resolution on the determination of temperature, gravity and metallicity are small. We proceed in two steps: 

\noindent
$(1)$ fits to the profiles of the hydrogen Balmer lines (H4 to H10) are used to fix \logg\ as a function of \teff. 
The decreasing equivalent width of the nebular Balmer lines with increasing line order makes the analysis possible even when moderate nebular contamination is present. The uncertainties in \logg\ are estimated from the results obtained for the different Balmer lines. Examples based on the final \teff\ adopted (from step $2$) are illustrated in Figure~\ref{fig:Balmer-fits}.

\noindent
$(2)$ minimizing the $\chi^2$ values calculated as a function of [$Z$] and \teff\ in selected wavelength windows, containing a large number of metal lines from different atomic species and ions (such as \feii, \mgii, \tiii\ and \crii),  provides metallicities and stellar temperatures, with the corresponding uncertainties. The uncertainties are obtained from $\Delta \chi^2$ isocontours in the metallicity-temperature plane. Detailed examples are shown in \citet{Urbaneja:2023} and many other previous presentations of our work. Figure~\ref{fig:metal-fits} shows the final fits for the same three stars included in Figure~\ref{fig:Balmer-fits}.

\subsection{Additional quantities}\label{sec:additional_quantities}
For the purpose of constraining the evolutionary status of the \bsg s that we have analyzed, as well as to derive a spectroscopic distance to M101 (Section~\ref{sec:distance}), we have calculated the following quantities:\\[-3mm]

\noindent
{\em (a) Color excess and visual extinction -- } The spectral energy distribution of the adopted stellar models provide theoretical $B-V$ and $V-I$ color indices, which, when compared with the observed HST colors, can be used to yield the color excesses due to interstellar absorption, \mbox{$E(B-V)$} and $E(V-I)$. A second value of $E(B-V)$, derived from the relation $E(B-V) = 0.75\,E(V-I)$ (obtained from the \citealt{Cardelli:1989} reddening law), is averaged with the first one to obtain the adopted $E(B-V)$. The visual interstellar extinction is then calculated as $A_V = R_V \cdot E(B-V)$, with $R_V = 3.3$. The $E(B-V)$ uncertainties result from the uncertainties in the observed photometry and the stellar parameters obtained in the spectroscopic analysis. The theoretical colors depend on temperature, gravity and metallicity, with the strongest dependence on \teff. For the final $E(B-V)$ errors we add the photometric and stellar parameter uncertainties in quadrature. We note that the accurate determination of reddening is crucial for the determination of distance and that the error in interstellar extinction is considered in the error estimate for the distance modulus.

The mean $E(B-V)$ color excess of the stars in Table~\ref{table:2} is $0.12 \pm 0.05$.
From the mean logarithmic extinction coefficient at \hbeta, $c(H\beta)$, of the \hii\ region sample observed by \citet{Croxall:2016}, considered in the analysis presented in Section~\ref{sec:gradient}, we obtain a mean $\langle E(B-V)\rangle = 0.11\pm0.07$\footnote{Since $c(H\beta) = 0.4\, R_{H\beta}\cdot E(B-V)$ (\eg\ \citealt{Aller:1984}), adopting the Cardelli reddening law with $R_V = 3.3$ and $R_{H\beta} = 3.81$, we obtain $E(B-V) = 0.66\,c(H\beta)$.}, in excellent agreement with the \bsg s.\\
\\[-2mm]
\noindent
{\em (b) Bolometric magnitude --} The adopted stellar models yield a value for the bolometric correction $BC_V$, which allows us to calculate the apparent bolometric magnitudes as $m_{bol} = V + BC_V$.\\[-2mm]

\begin{figure}[ht]
	\center \includegraphics[width=1\columnwidth]{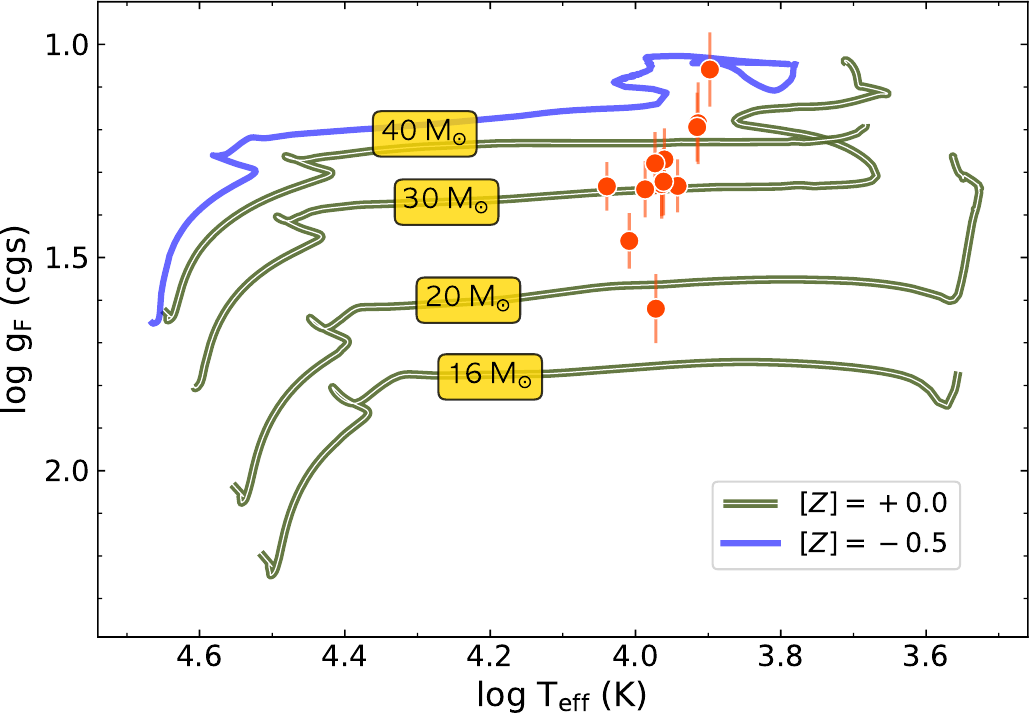}\medskip
	\caption{Spectroscopic Hertzsprung-Russell diagram of the blue supergiants in M101. We include MESA evolutionary tracks for [$Z$]\,=\,0.0 (initial stellar masses: 16, 20, 30 and 40\,\msun) and [$Z$]\,=\,$-0.5$ (initial stellar mass: 40\,\msun).}\label{fig:sHRD}
\end{figure}

\noindent
{\em (c) Flux-weighted gravity --} This quantity, defined as 
$\log g_F = \log g - 4 \log$(\teff/10$^4$\,K), can be used to constrain the evolutionary status of the \bsg s and derive their distance (\citealt{Kudritzki:2003}). We defer the presentation of the distance calculation to Section~\ref{sec:distance}, while here we analyze the evolutionary status of the supergiants, looking at the spectroscopic Hertzsprung-Russell diagram of Figure~\ref{fig:sHRD}. Such a diagram, where $\log$~\teff\ is shown against $\log g_F$, can be used to compare the observed stellar parameters with those predicted by stellar models in a distance-independent manner (\citealt{Langer:2014}). 

In Figure~\ref{fig:sHRD} we display MESA evolutionary tracks (\citealt{Choi:2016, Dotter:2016}) calculated for solar metallicity and including stellar rotation effects, in the range 16-40~\msun\ (green lines). Our analysis reveals a significant metallicity variation among our targets, from [$Z$]\,=\,$-0.53\pm0.06$ to [$Z$]\,=\,$0.29\pm0.06$. Since our aim is merely to provide an indicative mass range for our stellar targets, the adoption of the solar metallicity evolutionary tracks is sufficient, even though in Figure~\ref{fig:sHRD} we add  the 40~\msun\ track for [$Z$]\,=\,$-0.5$ (blue line), in order to better describe the status of star \#02, the object with the lowest $\log g_F$ value and the lowest metallicity in our sample.
As we already found in our previous spectroscopic studies of blue supergiants in nearby spiral galaxies (\eg\ \citealt{Bresolin:2022, Kudritzki:2024}), the evolutionary tracks indicate initial stellar masses of the \bsg s in the range 20-40~\msun, with ages ranging approximately between 5 and 10~Myr.

\section{The metallicity gradient} \label{sec:gradient}

We present here the first determination of the radial stellar metallicity gradient of M101, using the results from the previous section, and carry out a comparison with the well-studied nebular abundance gradient determined from \hii\ regions. In order to proceed, we make the usual assumptions that the nebular oxygen abundance represents the metallicity of the ionized gas and that the abundance pattern is solar, and adopt \eosun\ = \ohsun 8.69 from \citet[see also \citealt{Asplund:2009}]{Allende-Prieto:2001}.

For the comparison, we take the \hii\ region study of \cite{Croxall:2016}, who determined O/H and N/O abundances from ratios of collisionally excited lines (\cel s) including temperature-sensitive auroral lines for 74 nebulae using the so-called ``direct method''. The galactocentric radial 
gradient they obtained agrees with other similar measurements in the literature, based on smaller samples (\citealt{Kennicutt:2003, Bresolin:2007, Li:2013, Hu:2018, Esteban:2020}).

\begin{figure*}
	\center
	\includegraphics[width=0.8\textwidth]{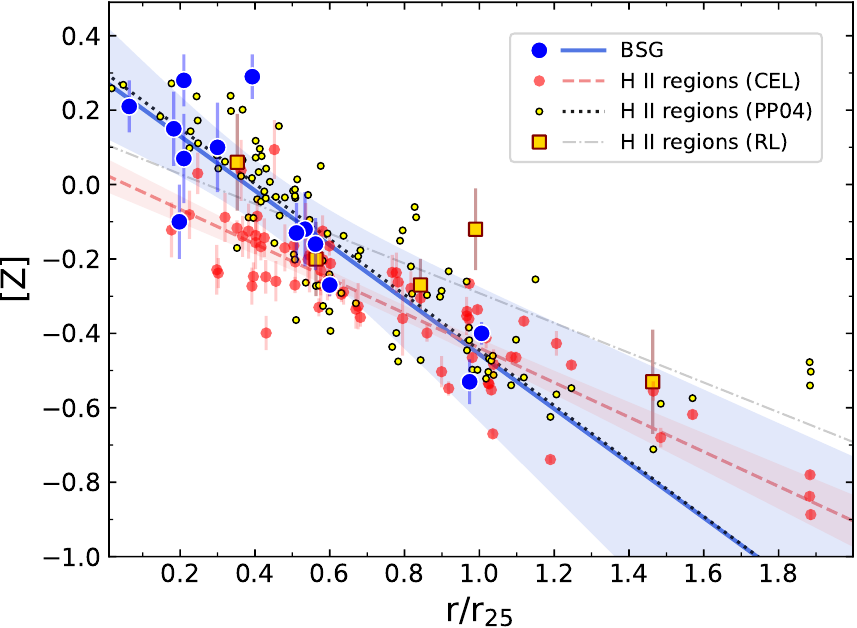}
	\caption{Galactocentric metallicity gradient in units of the isophotal radius \rtf, as determined from \bsg s (blue dots) and \hii\ regions using (a) \cel s and the direct method (\citealt{Croxall:2016}, red dots),  (b) the O3N2 strong-line index (\citealt{Pettini:2004}, yellow dots) and (c) \rl s (\citealt{Esteban:2020}, gold squares).
		Linear regressions to these four sets of data points are indicated by the continuous, dashed, dotted and dash-dotted lines, respectively. The shaded regions represent the 95\% confidence intervals of the linear fits to the stellar and direct nebular metallicities. The nebular abundances do not account for dust depletion effects. Doing so would increase the \hii\ region metallicities.
	}\label{fig:gradient}\bigskip
\end{figure*}

Figure~\ref{fig:gradient} illustrates the stellar (\bsg: blue symbols) and 
nebular (\hii\ regions with the direct method: red symbols) metallicity radial gradients. 
We carried out an orthogonal distance regression (using {\tt scipy.odr}), assigning a conservative error of 1 arcsec to the radial coordinate.
We obtain the following fits, also shown in Figure~\ref{fig:gradient}:
\begin{equation}\label{eq:gradient_stars}
\mathrm{stars:}~~					[Z] = 0.277~ (\pm 0.083) - 0.733~ (\pm 0.117)~ r/r_{25} 
\end{equation}	

\begin{equation}\label{eq:gradient_hii}
	\mathrm{H\,\text{\footnotesize II}:}~~				[Z] = 0.026~ (\pm 0.033) - 0.465~ (\pm 0.030)~ r/r_{25}. 
\end{equation}	

We note that Equation~(\ref{eq:gradient_hii}) differs from Equation~(10) in \citet{Croxall:2016}, expressed in terms of metallicity rather than oxygen abundance, because of the different choice of isophotal radius and regression algorithm.

Although formally the stellar gradient is steeper than the nebular gradient, the difference is at the $2\sigma$ level, and should not be considered statistically significant. 
The zero point discrepancy is mitigated and becomes non-significant if we account, as we have done in our previous investigations, for a nebular oxygen depletion into dust on the order of 0.1~dex (see \citealt{Bresolin:2016}).
By increasing the nebular metallicities by 0.13 dex we minimize the scatter of the stellar data in relation to the nebular regression line, and the zero points differ by $\sim$1.4$\sigma$. In Section~\ref{sec:broader} we will further discuss this issue.

A limiting factor in our comparison, and the reason for the larger errors in the stellar regression, is represented by the fact that the stars span a much narrower radial range compared to the  \hii\ regions, almost by a factor of two. On the other hand, it is  worth pointing out that our stellar abundances reach closer to the galactic center than the nebular abundances determined by \citet{Croxall:2016}: this is due to the scarcity of bright ionized nebulae suited for the analysis, as well as the increased difficulty in detecting the weak auroral lines at higher metallicities, while the stellar absorption features in our \bsg\ spectra become more prominent. Our data, therefore, provide direct evidence for a central metallicity in M101 of $\sim$1.9~\zsun.

In their study of the spiral galaxy M83, \citet{Bresolin:2016} made a comparison between stellar and nebular abundances, the latter calculated with a number of different strong-line abundance diagnostics. They reached the conclusion that the calibration of the diagnostic O3N2 = log\{(\oiii\lin5007/\hbeta)/(\nii\lin6583/\halpha)\} by \citet[hereafter PP04]{Pettini:2004} provides the best match between metallicities determined from blue supergiants and from \hii\ regions. Recently, \citet{Kudritzki:2024} found reasonable agreement between stellar and ionized gas metallicities using the same nebular method in NGC~4258. 
We include in Figure~\ref{fig:gradient} the \hii\ region metallicities that we derive from the full sample (102 objects) observed by \citet{Croxall:2016}, adopting the O3N2 calibration by PP04, including objects lacking auroral line detections (yellow dots). There is a remarkable agreement with our \bsg\ metallicities. A linear regression, limited to the same interval covered by the stars, is virtually indistinguishable from the stellar regression of Equation~(\ref{eq:gradient_stars}):
\begin{equation}\label{eq:gradient_PP04}
	\mathrm{O3N2:}~~					[Z] = 0.300~ (\pm 0.028) - 0.745~ (\pm 0.044)~ r/r_{25}.
\end{equation}

Although we noticed a similar agreement in our previous work, this is still a somewhat surprising result, given the relatively small number of calibrators available to PP04 compared to more recent strong-line diagnostic calibrations (\eg\ \citealt{Pilyugin:2016}), the omission of depletion of oxygen onto dust grains in our comparison, and the fact that the PP04 relation greatly depends on photoionization models of four \hii\ regions at supersolar oxygen abundance. Note that in Figure~\ref{fig:gradient} the divergence between the linear fit and the points corresponding to objects in the outer disk is due to the flattening of the metallicity with increasing O3N2 (see PP04). 

Carrying out a deeper investigation into the results from a selection of strong-line abundance indicators, as done by \citet{Bresolin:2016}, is beyond the scope of our study. However, we verified that the S calibration by \citet{Pilyugin:2016}, based on direct abundances of a considerably larger sample of \hii\ regions compared to the PP04 diagnostic, provides abundances that are in broad agreement with those obtained from the direct method.

Figure~\ref{fig:gradient} also includes the five \hii\ regions whose gas-phase oxygen abundances were obtained by \citet{Esteban:2009, Esteban:2020} from the detection of \oiiION\ recombination lines (\rl s).
Because of the known abundance discrepancy in \hii\ regions (\citealt{Garcia-Rojas:2007}), \rl-based abundances are systematically offset to $\sim$0.15 higher values compared to \cel-based measurements. Additional discussion on these \rl\ data in relation to the \bsg\ data will appear in Section~\ref{sec:broader}.

\begin{figure*}
	\center
	\includegraphics[width=0.8\textwidth]{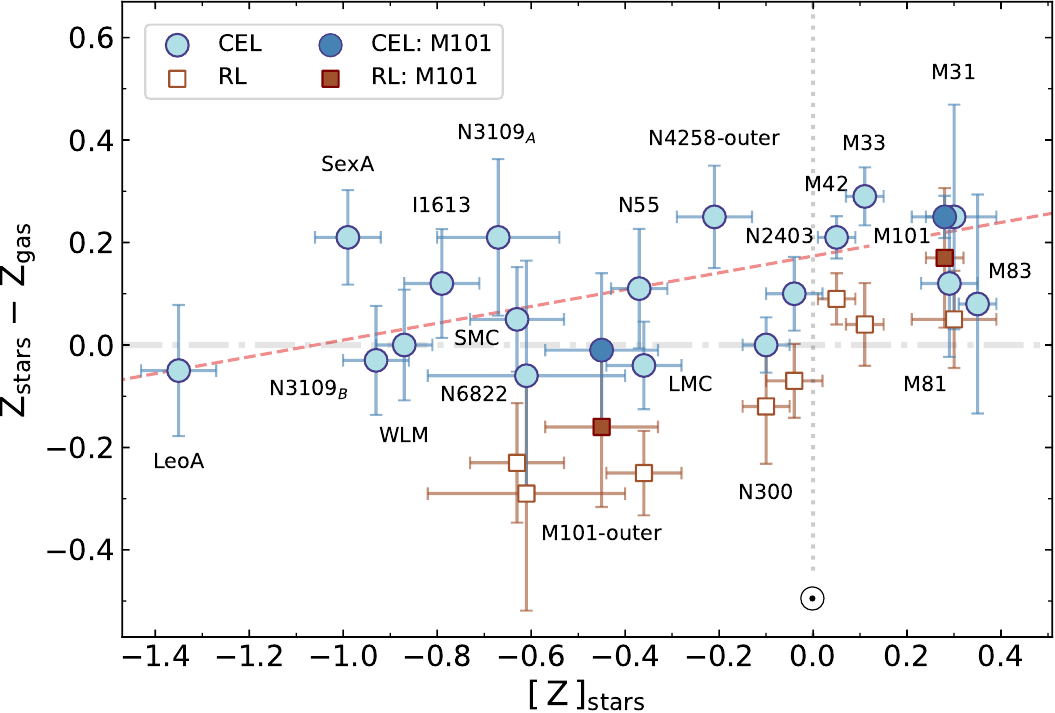}
	\caption{Metallicity difference between \bsg s and \hii\ regions as a function of stellar metallicity in a sample of 18 star-forming galaxies. The nebular values are determined from collisionally excited lines (\cel s: blue circles), using the direct method, and metal recombination lines (\rl s). For all spiral galaxies in the plot, the metallicities are central values (galactocentric distances $r$\,=\,0) obtained from the regressions to the logarithmic abundances $[Z]$. For M101 we also show an additional data point at $r$\,=\,\rtf\ (M101-outer). For the irregular galaxies a mean abundance is adopted.
    The dashed line represents the linear regression to the \cel-based data points, discussed in Section~\ref{sec:dust_depletion}.
    The nebular abundances do not account for dust depletion effects. Doing so would reduce the values in the abscissa.
	}\label{fig:stars_vs_hii}\bigskip
\end{figure*}

\section{The broader context}\label{sec:broader}
Among the main objectives of our series of papers about the \bsg s of external galaxies is the comparative study of the metallicity of the resolved young populations that can be carried out using both massive stars and the ionized gas to trace the abundance of metals (see \citealt{Bresolin:2022, Urbaneja:2023, Kudritzki:2024} for recent results). A related, but independent, line of investigation focuses on the chemical abundances of unresolved stellar populations, based on the analysis of the absorption lines of their composite spectra. This alternative avenue, applied to the young population, has been producing results that are in good agreement with those obtained from the study of individual stars, and promises to facilitate the {\em stars vs ionized gas} chemical abundance comparison at larger cosmic distances and with a smaller investment of telescope time (\citealt{Sextl:2024}).

The techniques used to derive the metal content of \bsg s and \hii\ regions are notoriously vastly different, relying on the analysis of either absorption or emission lines, respectively, and the physical processes that determine these spectral features are totally distinct. For decades the analysis of the nebular spectra of extragalactic \hii\ regions has represented the primary technique used to measure the chemical abundances of star-forming galaxies, thanks to the relative simplicity of the observations and analysis required, as well as the high intrinsic emission line luminosity of the target ionized nebulae.
Yet, the abundance scale based on the observation of emission lines from \hii\ regions is unfortunately plagued by large systematic uncertainties, that become apparent when different abundance measurement methods are employed (\citealt{Moustakas:2010, Lopez-Sanchez:2012, Bresolin:2016}). 
An extensive comparison between stellar and nebular metallicities in galaxies is crucial in order to investigate this issue, because the systematic uncertainties affecting the chemical analysis of massive, blue supergiant stars are considered to be much smaller than for \hii\ regions. 

In this section we consider the metallicity of the \bsg s in M101 within the broader context of 
similar studies we have carried out in a number of additional galaxies, starting with a close look at the comparison with the nebular abundances.

\subsection{Stellar versus nebular metallicities}\label{sec:gas_vs_stars}

We present the current status of our ongoing investigation on the comparison between stellar and nebular metallicities by illustrating in Figure~\ref{fig:stars_vs_hii} the difference 
$\Delta_Z$ = $Z_{stars} - Z_{gas}$
between \bsg\ and \hii\ region metallicities as a function of stellar metallicities, updating our previous results (see \citealt{Bresolin:2016} for literature references, with the addition of the more recent studies by \citealt{Bresolin:2022}, \citealt{Urbaneja:2023} and \citealt{Kudritzki:2024}). The nebular metallicities are scaled with the O/H abundances, adopting \ohsun\, 8.69 (\citealt{Asplund:2009}), and are determined either by the direct method (using \cel s: blue circle symbols in the diagram) or \oiiION\  recombination lines (\rl s: square symbols). 
We have updated the gas-phase abundances of the LMC and the SMC, using weighted averages of the individual \hii\ region oxygen abundances presented by \citet[\cel s]{Dominguez-Guzman:2022} and \citet[\rl s]{Toribio-San-Cipriano:2017}.

For irregular galaxies, characterized by absent or minimal radial abundance gradients, we adopt mean abundances. For spirals we adopt the central values that can be inferred from a linear fit to their logarithmic galactocentric abundance gradients.
In the case of M101, we take advantage of the significant radial metallicity gradient obtained from \bsg s, nebular \cel s (\citealt{Croxall:2016}) and nebular \rl s (\citealt{Esteban:2009, Esteban:2020}). We calculate two values: one corresponding to the linear regression intercept ($r$\,=\,0) and one at $r$\,=\,\rtf\ ('M101-outer' in the diagram), because at this radial distance the linear fits are well constrained by the presence of \bsg s (see Figure~\ref{fig:gradient}), as well as \hii\ regions with \rl\ data (\citealt{Esteban:2020}). Thus, in Figure~\ref{fig:stars_vs_hii} we show two points that refer to M101, spanning the range of metallicities that we observe for the \bsg s in this galaxy.

In our previous work we have remarked about the broad agreement, over a considerable range in metallicity (1.7 dex, a factor of 50), between \bsg s and \hii\ regions when the nebular abundances are determined from \cel s, once a relatively small dust depletion factor for oxygen is introduced to increase the measured gas-phase metallicities. We attribute the weighted difference for the data points shown in Figure~\ref{fig:stars_vs_hii}, $\Delta_Z = 0.151\pm0.018$ dex, to such a depletion of oxygen. In Section~\ref{sec:dust_depletion} we will consider the metallicity dependence of $\Delta_Z$.

The gas metallicities obtained from \rl s (without considering depletion effects) are in good agreement with the stellar metallicities around the solar abundance, but the two appear to diverge with decreasing stellar metallicity, gas abundances being larger than stellar ones. 
There is a striking offset between stellar and \rl-based metallicities for the LMC and the SMC, arguably among the best studied galaxies in Figure~\ref{fig:stars_vs_hii} in terms of chemical composition. With the exceptions of the discussions by \citet{Bresolin:2016} and \citet{Toribio-San-Cipriano:2017}, this appears to be a crucially neglected problem, since the fact that \cel s appear to underestimate stellar abundances in the Milky Way is interpreted as supporting evidence for the presence of temperature fluctuations in \hii\ regions (\eg\ \citealt{Garcia-Rojas:2014}), which in turn are considered to provide a likely explanation for the systematic discrepancy the the oxygen abundances derived from \rl s and \cel s.
Even without considering the effect of dust depletion, \rl s overestimate the B star abundances in the LMC/SMC by $\sim$0.25 dex, which is of the same size (but opposite in sign) as the discrepancy factor observed in the Milky Way (\citealt{Garcia-Rojas:2007}).

Our investigation of the \bsg\ metal content in nearby resolved galaxies indicates that the direct method yields reliable gas metallicities, or at least that these two approaches of measuring metallicities in star-forming galaxies are consistent with each other.
This can be seen as a puzzling result, because there is convincing evidence that the presence of temperature inhomogeneities in \hii\ regions cause the analysis of \cel s to underestimate the gas abundances (\citealt{Mendez-Delgado:2023}). We are unaware of the existence of processes that would make the analysis of the spectra of massive stars replicate the effects that temperature fluctuations have on the derivation of chemical abundances from the analysis of nebular spectra.

\subsubsection{Estimating the dust depletion of oxygen in photoionized regions}\label{sec:dust_depletion}

In our comparison, we need to account for the mild depletion of oxygen atoms onto dust grains in the interstellar medium (\citealt{Jenkins:2009, Psaradaki:2023, Konstantopoulou:2024}). In the past we adopted a fixed value of 0.1 dex, in broad agreement with recent studies of \hii\ regions and star-forming galaxies that make use of photoionization models (\eg\ \citealt{Dopita:2013, Gutkin:2016, Amayo:2021}). For simplicity, we have insofar not accounted for the metallicity dependence of dust depletion (\citealt{Roman-Duval:2022, Hamanowicz:2024}).

We attribute the offset $\Delta_Z$ between stellar and nebular (\cel) metallicities  that we observe for the sample of galaxies in Figure~\ref{fig:stars_vs_hii} exclusively to the depletion of oxygen from the gas phase to the dust phase. 
From Section~\ref{sec:gas_vs_stars}, the weighted difference is $\langle\Delta_Z\rangle\, =\,0.151 \pm 0.018$ dex (standard deviation: 0.113 dex). Thus, we can to first order conclude that
a fixed oxygen depletion factor $\delta$(O)\,$\simeq$\,0.15 dex would bring our stellar and nebular (\cel) data in good agreement with each other.\footnote{In studies of the neutral ISM the logarithmic depletion factor $\delta$ is defined as a negative quantity  (\eg\ \citealt{Galliano:2018}). In \hii\ region work it is customary to discuss it with a reversed sign, as done here.}

Next we consider the possibility that $\delta$(O) varies with metallicity.
From an analysis of the increasing Ne/O abundance ratio with O/H for extragalactic \hii\ regions, \citet{Amayo:2021} estimated an increasing trend for $\delta$(O) from zero to 0.12 dex in the range \oh\,=\,(7.0, 8.67), equivalent to [$Z$]\,=\,($-1.69$, 0.02). If we adopt such a relation, the standard deviation of the weighted difference $\Delta_Z$ decreases to 0.099, suggesting that such a solution should be preferred over a constant $\delta$(O) value.

Performing an orthogonal distance regression fit to the $\Delta_Z$ data in our sample of galaxies we obtain:

\begin{equation}\label{eq:depletion_factor}
\delta({\mathrm O}) = 0.174~ (0.023) + 0.164~ (0.052)~ [Z],
\end{equation}

\noindent
with a standard deviation of 0.092 of the weighted difference $\Delta_Z$.

This relation, shown as a dashed line in Figure~\ref{fig:stars_vs_hii}, is steeper than the relation implied by the study by \citet[slope\,=\,0.07]{Amayo:2021}, and reaches $\delta$(O)\,=\,0 at [$Z$]\,=\,$-1.06$ (\oh\,=\,7.63).
From this analysis we can conclude that below this metallicity, about 0.1\,$\times$\,solar, the dust depletion of oxygen in the ionized gas of the sample galaxies is negligible, and that it increases to $\sim$$0.17\pm0.02$ dex at solar metallicity. Our result does not change significantly if we remove the most metal-poor galaxy in our sample, Leo~A, from the analysis.
The oxygen depletion level we deduce around the solar metallicity is consistent with the determination in the Orion nebula by \citet{Mesa-Delgado:2009}, who estimated an average value, from three different approaches, of $\sim$$0.20\pm0.03$ dex if relying only on \cel s (\ie\ neglecting temperature fluctuations).

\begin{figure}
	\center
	\includegraphics[width=1\columnwidth]{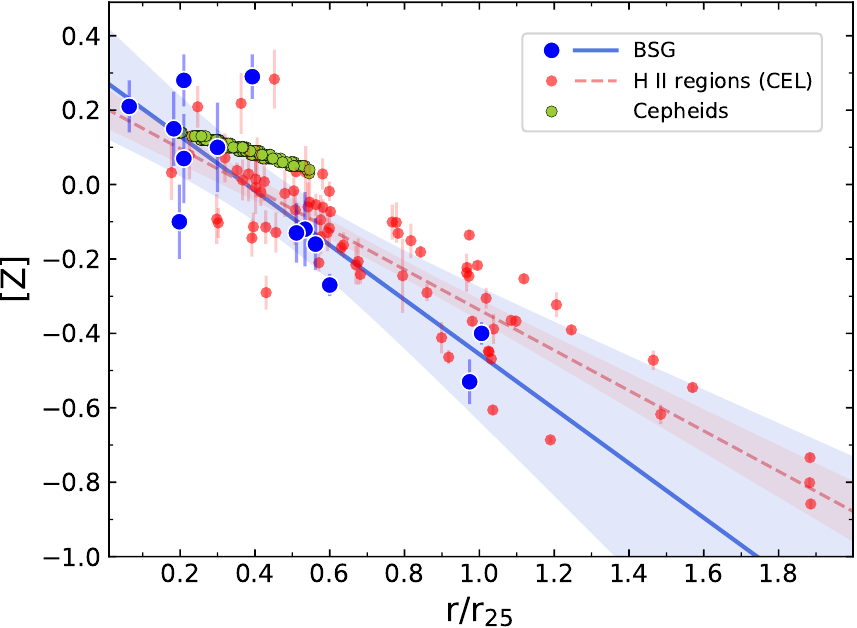}
	\caption{As Figure~\ref{fig:gradient}, but applying the dust depletion factor in Equation~(\ref{eq:depletion_factor}) to the gas-phase metallicities. For clarity, we removed the data points corresponding to \hii\ regions with metallicities from the PP04 method and those from the use of recombination lines. The green data points correspond to the metallicities of M101 Cepheids as adopted by \citet{Riess:2022} in their determination of the Hubble constant (for a detailed discussion, see Section \ref{sec:cepheids}).
	}\label{fig:gradient2}\bigskip
\end{figure}

As a test of our procedure, in Figure~\ref{fig:gradient2} we reproduce the \bsg\ and \hii\ region metallicity gradients in M101 shown in Figure~\ref{fig:gradient}, but applying $\delta$(O), calculated from Equation~(\ref{eq:depletion_factor}), to the gas-phase metallicities.
The figure illustrates the good agreement between stellar and direct nebular metallicities that we obtain by taking a metallicity-dependent dust depletion factor for oxygen into account.
A corollary of the dependence of dust depletion on metallicity is that the actual radial oxygen abundance gradients of spiral galaxies are steeper than those measured from \hii\ regions. In the case of M101, shown in Figures~\ref{fig:gradient} and~\ref{fig:gradient2}, and applying Equation~(\ref{eq:depletion_factor}), the slope increases by 17\%, from $0.465~ r/r_{25}$ to $0.542~ r/r_{25}$.

In our analysis of the depletion of oxygen in the photoionized gas of a sample of galaxies, we are assuming that there are no significant variations between galaxies, apart from the metallicity effect.
While it is well-known that the amount of metal depletion on dust grains measured in the neutral ISM of the Milky Way varies widely with sightline (\eg\ \citealt{Savage:1979}), it is not yet known whether the same applies to \hii\ regions. Studies are under way to utilize the depletion scheme introduced by \citet{Jenkins:2009}, that includes a parameter ($F_\ast$, in the range [0,1]) that measures the degree of metal depletion, common to all elements, on observations and modeling of photoionized regions, and test whether a small range in $F_\ast$ values is appropriate (\citealt{Gunasekera:2022, Gunasekera:2023}).
In any case, the slope of the relation between $\delta({\mathrm O})$ and $F_\ast$ is rather 
shallow ($0.23$; \citealt{Jenkins:2009}), implying that significant variations in $F_\ast$ can still result in small changes in  $\delta({\mathrm O})$.
Indirect evidence for a restricted range in the depletion of oxygen in photoionized regions is offered by the typically small abundance dispersions, $\lesssim$\,0.1 dex, about the radial abundance gradients determined for spiral galaxies from the analysis of their \hii\ regions (\citealt{Bresolin:2011, Croxall:2016, Rogers:2022}), and the chemical homogeneity of dwarf galaxies as determined from \hii\ regions  (\citealt{Lee:2006b, Dominguez-Guzman:2022}).
An additional clue is given by the rather small range in 
the extinction coefficients measured for extragalactic \hii\ regions, $c(H\beta) \simeq$ 0.1-0.4, which can be associated to an equivalently narrow range in the hydrogen column density $N$(H) along the line of sight. Since the amount of metals locked in dust increases with $N$(H) (\citealt{Roman-Duval:2022}), which reflects the increase of depletion with mean line of sight density (\citealt{Savage:1979a, Jenkins:2009}), it appears reasonable to expect relatively small variations in $\delta({\mathrm O})$.

While our model of dust depletion seems compelling, a critical remark at the end of this subsection is in order. One can argue that with Equation~(\ref{eq:depletion_factor}) we try to interpret a relatively small systematic discrepancy between stellar and nebular diagnostics, which could have an entirely different origin. One obvious alternative could be a systematic variation of the $\alpha$/Fe element ratio with metallicity and the fact our \bsg\ spectral analysis adopts solar $\alpha$/Fe and yields metallicities dominated by iron group elements (see \citealt{Bresolin:2022}). Spectroscopy of young, massive stars in the Milky Way and the Magellanic Clouds  does not indicate systematic deviations from solar $\alpha$/Fe, but the situation could be different in other galaxies. However, generally one would expect an increase of $\alpha$/Fe at low metallicity, which is just the opposite to our Equation~(\ref{eq:depletion_factor}).

\subsection{A comparison with the SH0ES project}\label{sec:cepheids}

M101 is one of the 37 galaxies in the SH0ES project, which aims at a precision determination of the Hubble constant ($H_0$) through distance determinations with Cepheid stars and the subsequent use of Type Ia supernovae (SNe~Ia; \citealt{Riess:2022}). Since the zero-point  of the Cepheid period-luminosity relation (PLR) depends on metallicity, it is important to compare our \bsg\ metallicities with those used by Riess et al. in their derivation of Cepheid-based distances. Both \bsg s and Cepheids are types of massive stars  belonging to the same young population. However, Cepheids are substantially fainter than \bsg s and, thus, at the distance of the SH0ES galaxies, a direct spectroscopic determination of their metallicities is not possible. Therefore, the SH0ES project has to rely on \hii\ region metallicities as a proxy. \citet{Riess:2022} apply a strong line method to obtain these metallicities but, in view of the uncertainties of the individual strong line methods, they do not rely on one single calibration, using instead an average of nine calibrations as provided by \citet{Teimoorinia:2021}. They point out that their method yields results in very close agreement with the PP04 method.

Table 2 in \citet{Riess:2022} contains the metallicities adopted for all individual Cepheids in the SH0ES project, together with their coordinates. This allows us to compare these metallicities with those obtained in our work. The comparison is carried out in Figure~\ref{fig:gradient2}, where green dots represent the Cepheid stars.
Although these lie at a systematically higher metallicity compared to \bsg s,
the difference is very small. Compared with the \bsg\ regression, the mean difference in metallicity is 0.08 dex. With a metallicity dependence of the PLR zero-point magnitude $\Delta m = -0.217[Z]$ mag/dex (see \citealt{Riess:2022}), a correction of the Cepheid metallicities to the \bsg\ values would result in a distance modulus shorter by 0.017 mag. This is clearly less than the distance modulus uncertainty for M101 of 0.04 mag given in \citet{Riess:2022} and does not have any serious effect on the determination of $H_0$. We also note that in our \bsg\ study of the Hubble constant anchor galaxy NGC~4258 we have found a very similar result (\citealt{Kudritzki:2024}). Our \bsg\ metallicities are in close agreement with those used by Riess et al. for the Cepheids. While the statistical experiments carried out in \citet{Riess:2022} have already shown that the effect of Cepheid metallicity on the determination of $H_0$ is small, the fact that we find this agreement is reassuring.  

\subsection{The galaxy mass-metallicity relation}\label{sec:MZR}

The scaling relation existing between galaxy stellar mass and average metallicity is known from studies of both stellar and gas-phase metal abundances (\citealt{Lequeux:1979, Tremonti:2004, Gallazzi:2005, Zahid:2017, Sextl:2023}).
This mass-metallicity relation (MZR) can be understood in terms of the interplay between gas infall, loss of metals due to feedback processes and downsizing in regulating the chemical evolution of galaxies (see \citealt{Maiolino:2019} for a recent review).

As we have done in previous publications, we present here our new results -- for M101 in this case -- within the context of
the MZR based purely on the metallicity of the young stellar populations obtained from the spectral analysis of absorption lines. Over the course of the past two decades our work has been based on the spectral analysis of \bsg s, red supergiants (\rsg s) and super star clusters (\ssc s), as well as population synthesis modeling. References can be found in \citealt{Bresolin:2022}, with the addition of the analysis of Leo~A by \citet{Urbaneja:2023}, NGC~4258 (\citealt{Kudritzki:2024}), NGC~1365 (\citealt{Sextl:2024}) and M83 (\citealt{Sextl:2025}).

In the case of M101, we adopt a stellar mass $\log M_\ast/$\msun~=~10.35 from \citet{Leroy:2019}, after a correction for the different distance we use, and, for consistency with our previous determinations, the stellar metallicity we infer from the galactocentric radial gradient in Equation~(\ref{eq:gradient_stars}) at $r=0.4$~\rtf.
This choice is motivated by studies of \hii\ region O/H abundances, such as those by \citet{Zaritsky:1994} and \citet{Moustakas:2006}, that indicate that the chemical abundance measured at such a distance from the center is representative of the global value. 

\begin{figure}
	\center
	\includegraphics[width=1\columnwidth]{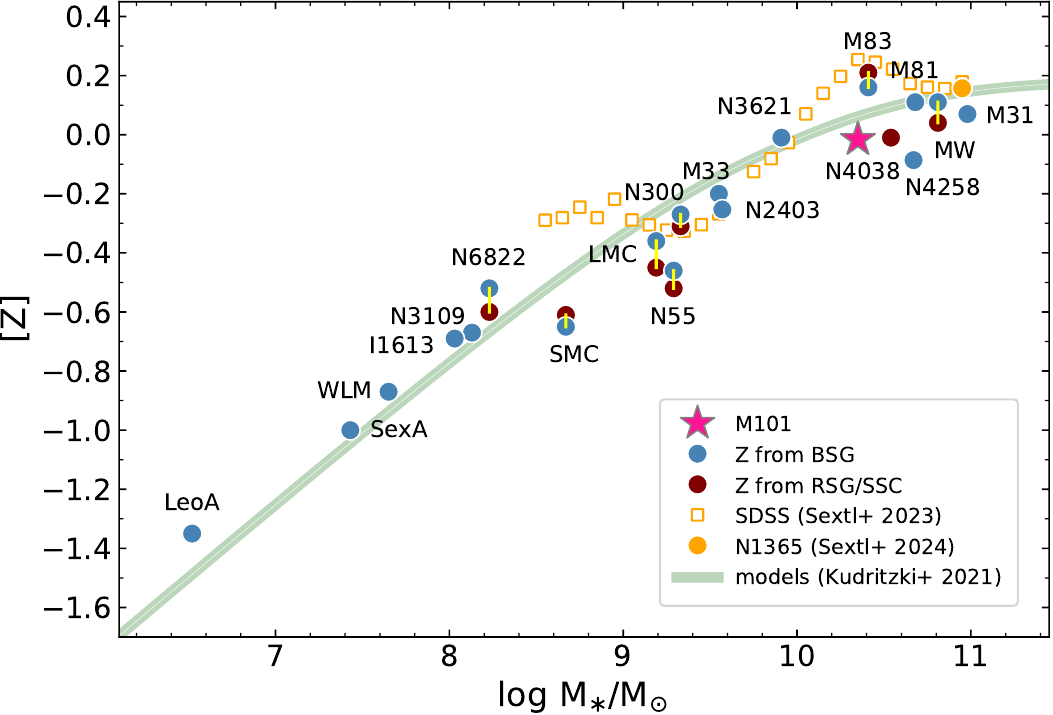}
	\caption{Stellar mass--metallicity relation based on investigations of the young stellar populations of star-forming galaxies from the spectral analysis of absorption lines, including M101 (pink star symbol). The legend and the text explain the meaning of the data illustrated by the diagram. 
	}\label{fig:mzr}\bigskip
\end{figure}

The updated MZR, including a data point for M101, extends over approximately four orders of magnitude in stellar mass, and is shown in Figure~\ref{fig:mzr}, where different symbols are used for metallicity determinations made from \bsg s (blue dots), red supergiants (\rsg s) and super star clusters (\ssc s) (red dots), and population synthesis (orange dot).
We note that the population synthesis analysis of M83 by \citet{Sextl:2025} yields a very flat metallicity distribution of the young stellar population in the inner part of the galaxy. From their galactocentric radial distribution we obtain, at $r=0.4$~\rtf, a value of $[Z] = 0.16$, which is in excellent agreement with the two data points (obtained from the spectral analysis of \bsg s\ and \rsg s, respectively) shown for this galaxy.
Figure~\ref{fig:mzr} also includes the results from population synthesis analysis of 250,000 star-forming Sloan Digital Sky Survey galaxies (\citealt[orange squares]{Sextl:2023}) and predictions from the galaxy evolution models by \citet[green curve]{Kudritzki:2021a, Kudritzki:2021b}. The M101 data point fits well into the the upper, flattening part of the MZR.

Finally, a word of caution: comparisons between the MZR derived from gas phase metallicities (O/H) and from stellar abundances should account for the unseen fraction of oxygen atoms locked in interstellar dust grains, as done in Section~\ref{sec:gas_vs_stars}. An additional systematic offset can be introduced by 
referring the metallicity to different radii (\eg\ galaxy center \vs\ a fixed fraction of the isophotal radius).

\subsection{Metallicity gradients}

\begin{figure}[b]
	\center \includegraphics[width=1\columnwidth]{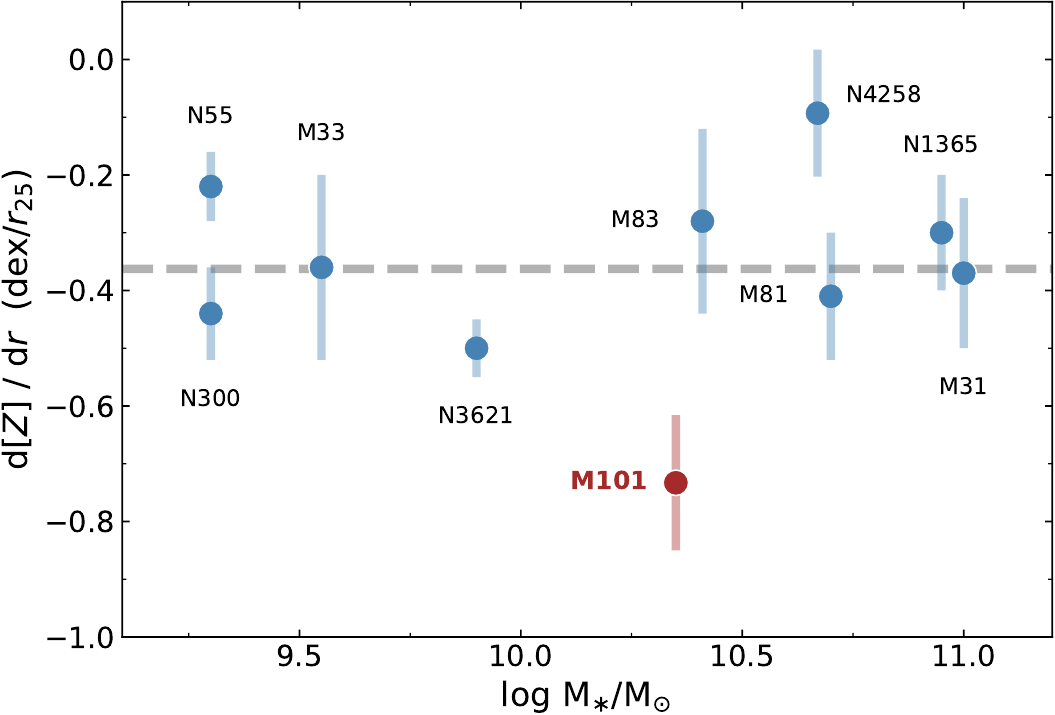}\medskip
	\caption{Comparison of metallicity gradients (in dex/\rtf) in spiral galaxies  as a function of galaxy stellar mass. The data have been derived in our \bsg\ spectroscopy project. The gradient of M101 is indicated in red. The data point for NGC~1365 results from the population synthesis study by \citet{Sextl:2024}. The horizontal line represents the weighted mean for the full sample, excluding M101.}\label{fig:r25}
\end{figure}

Metallicity gradients in star-forming disk galaxies provide important information to constrain the effects of matter accretion and galactic winds on the evolution of galaxies (see, for instance, \citealt{Kudritzki:2015, Ho:2015, Bresolin:2016, Kang:2023}). Compared with the results of our previous spectroscopic work on young stars in other galaxies, the gradient in M101, $-0.733\pm0.117$~dex/\rtf, is unusually steep. This can be seen in Figure~\ref{fig:r25}, which displays, as a function of galaxy stellar mass, the metallicity gradients of all galaxies that have been measured from studies of \bsg s, \rsg s and \ssc s. The value for NGC~1365 has been determined through a population synthesis study and represents the gradient of the young stellar population (see \citealt{Sextl:2024}, with the improvement described in \citealt{Sextl:2025}). M101 stands out as a galaxy with a very steep gradient.

\citet{Ho:2015}, in their \hii\ region study of metallicity gradients, have found average values of $-0.39 \pm 0.18$ or $-0.32\pm 0.2$ dex/\rtf\ (the error quoted is the standard deviation), depending on the galaxy samples selected.   The weighted mean slope in our sample of galaxies, excluding M101, is $-0.36 \pm 0.13$ dex/\rtf\ (horizontal line in Figure~\ref{fig:r25}). Compared with these values, M101 represents a 2$\sigma$ outlier.

We note that, adopting \rtf\,=\,14.4~arcmin from \citet{de-Vaucouleurs:1991} -- as done by many authors -- instead of \rtf\,=\,8.0~arcmin from \citet{Mihos:2013} (Sec.~\ref{subsec:selection}), would bring M101 in agreement with the remaining galaxies in Figure~\ref{fig:r25}. However, \citet{Mihos:2013} provide convincing arguments, supported by deep surface brightness photometry, that \citet{de-Vaucouleurs:1991} substantially overestimated the isophotal radius of this galaxy.
We plan a follow-up study of the evolution of M101, where we will investigate star formation and disk growth history using the methods described in \citet{Kang:2023} and \citet{Sextl:2024}.

\section{The spectroscopic distance to M101}\label{sec:distance}

The flux-weighted gravity, introduced in Section~\ref{sec:additional_quantities}, can be used to derive extragalactic distances by means of the flux-weighted gravity--luminosity relationship (\fglr) of blue supergiant stars, as demonstrated by \citet{Kudritzki:2003} and \citet{Kudritzki:2008}. This tool provides distance estimates that are in good agreement with other stellar distance indicators, such as Cepheids and the Tip of the Red Giant Branch (\trgb). The \fglr\ has been applied successfully out to the distance of the galaxy NGC~4258 ($D=7.5$~Mpc), where excellent agreement has been found by \citet{Kudritzki:2024} with the geometric distance obtained from the analysis of the nuclear maser.

We adopt the \fglr\ calibration by \citet{Urbaneja:2017}, based on observations of \bsg s in the Large Magellanic Cloud (LMC), updated to reflect the geometric distance to the LMC by \citet[see \citealt{Bresolin:2022, Kudritzki:2024}]{Pietrzynski:2019}:

\begin{equation}\label{eq:fglr1}
\log\,g_F \geq 1.30:~~~~	M_{bol} = 3.20\, (\log\,g_F - 1.3) - 8.518
\end{equation}
\begin{equation}\label{eq:fglr2}
\log\,g_F \leq 1.30:~~~~	M_{bol} = 8.34\, (\log\,g_F - 1.3) - 8.518.
\end{equation}

The uncertainty of the zero point in these two equations is 0.02 mag \citep{Urbaneja:2017}. This value is small compared to the distance modulus uncertainty obtained below.

\begin{figure}
	\center \includegraphics[width=1\columnwidth]{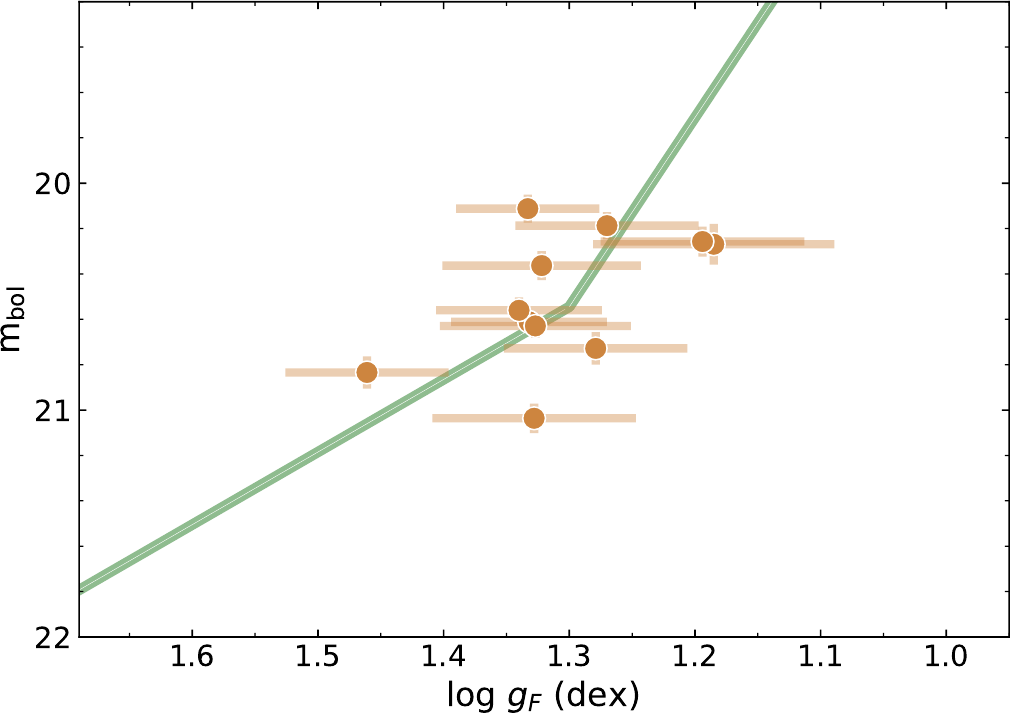}\medskip
	\caption{The \fglr\ in M101, showing the fiducial relation given by Equations~(\ref{eq:fglr1})-(\ref{eq:fglr2}) (green line), shifted along the vertical direction and fitted to the observed data points. The vertical shift provides the distance modulus to the galaxy.}\label{fig:fglr}
\end{figure}

Utilizing the apparent bolometric magnitudes and the \loggf\ values from Table~\ref{table:2}, we display in Figure~\ref{fig:fglr} the \fglr\ delineated by the \bsg s we have analyzed in M101, excluding \#2 and \#9, which deviate from the trend traced by the other targets by more than 2$\sigma$. In our previous work we have found that a fraction (10\%-15\%) of our spectroscopic targets follow the same behavior, which we attribute to unresolved stellar blends affecting the determination of magnitudes and stellar parameters.

We determine the vertical shift of the fiducial \fglr, represented by Equations~(\ref{eq:fglr1}) and (\ref{eq:fglr2}), that best fits the data points in Figure~\ref{fig:fglr}, as shown by the green line, using an 
orthogonal distance regression fit. This procedure yields a distance modulus 

$$\mu = m-M = 29.06 \pm 0.08$$

\noindent
corresponding to a distance of $6.5\pm0.2$~Mpc.

Our measurement of the distance to M101 is in good agreement with recent results obtained from primary distance indicators. The \trgb\ distance modulus from the online Extragalactic Distance Database\footnote{\href{https://edd.ifa.hawaii.edu}{https://edd.ifa.hawaii.edu}} (\citealt{Tully:2009}), based on HST imaging, is $\mu = 29.12\pm0.08$. 
This is similar to the Carnegie-Chicago Hubble Program values from HST observations, $\mu = 29.07\pm0.06$ (\citealt{Beaton:2019}) and from James Webb Space Telescope (JWST) observations, $\mu = 29.15\pm0.04$ (\citealt{Freedman:2025, Hoyt:2025}).

The most recent HST-based SH0ES distance modulus from Cepheids is $\mu = 29.18\pm0.04$ (\citealt{Riess:2022}), while Cepheids observed with JWST yield $\mu = 29.12\pm0.03$ (\citealt{Riess:2024}). \citet{Huang:2024} measured $\mu = 29.10\pm0.06$ from Mira variables, and \citet{Freedman:2025} obtained $\mu = 29.21\pm0.04$ from J-Region Asymptotic Giant Branch stars. The overall conclusion is that our spectroscopic distance is at the lower bound of recent determinations from other techniques, but still lying within $\sim$1$\sigma$ from measurements based on both \trgb\ and Cepheids.

\begin{figure}[ht]
	\center \includegraphics[width=1\columnwidth]{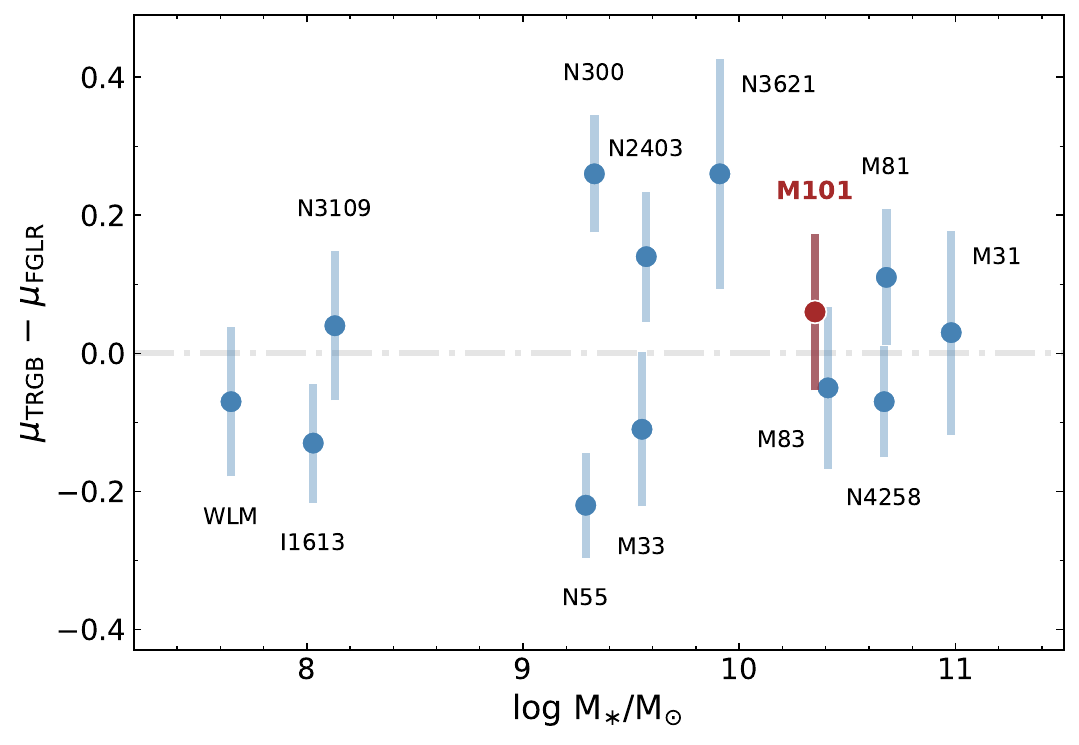}\medskip
	\caption{Difference in distance modulus between \trgb\ and \fglr\ measurements as a function of galaxy stellar mass for 13 galaxies.}\label{fig:trgb-fglr}
\end{figure}

We display in Figure~\ref{fig:trgb-fglr} the difference in distance modulus between the \fglr\ determinations we have carried out so far and those from the \trgb\ method (taken from \citealt{Tully:2009} and \citealt{Anand:2021}), as a function of galaxy stellar mass (see Section \ref{sec:MZR} for references regarding stellar mass). There are 13 galaxies shown, including M101 (red point), and their distance moduli are summarized in Table~\ref{table:FGLR_TRGB}. All \fglr\ distances are calculated with an orthogonal distance regression fit, as in the case of M101. This is different from our previous work (including \citealt{Kudritzki:2024} on NGC~4258), where we have used a simple weighted mean of the distances of the individual \bsg s in each galaxy with weights containing the contributions of the uncertainties of bolometric magnitude and flux-weighted gravity. Although the differences are small in most cases (with an average difference of $-0.05$ mag), the orthogonal distance regression is more appropriate. 

We do not find any statistically significant trend of the difference $\mu_{TRGB} - \mu_{FGLR}$, plotted along the y-axis, with galaxy stellar mass. There is also no significant vertical offset within the sample included in Figure~\ref{fig:trgb-fglr}: the weighted mean of the difference $\mu_{TRGB} - \mu_{FGLR}$ is $\Delta\mu = 0.00 \pm 0.03$, with a standard deviation of 0.15 mag.

\begin{deluxetable}{lccc}
	\tabletypesize{\footnotesize}	
	\tablecolumns{3}
	\tablewidth{0pt}
	\tablecaption{\fglr\ and \trgb\ distance moduli\label{table:FGLR_TRGB}}
	
	\tablehead{
		\colhead{Galaxy\phantom{aaaaaaaa}}	     		&
		\multicolumn{3}{c}{Distance Modulus}	\\
		\cline{2-4}
		\colhead{}       			&
		\colhead{\fglr}       		&
		\colhead{References}		&
		\colhead{\trgb}   }    		
	\startdata
	\\[-10pt]
	IC 1613    	&   24.52 $\pm$ 0.05 &  a   &   24.39 $^{+0.07}_{-0.04}$   	\\[4pt] 
	M31			&	24.55 $\pm$ 0.07 &  b   &   24.58 $^{+0.06}_{-0.13}$   	\\[4pt] 
	M33        	&   24.96 $\pm$ 0.05 &  b   &   24.85 $^{+0.10}_{-0.05}$  	\\[4pt]
	WLM			&	25.03 $\pm$	0.06 &  c	&	24.96 $^{+0.05}_{-0.09}$	\\[4pt]
	NGC 3109	&	25.59 $\pm$	0.04 &  d	&	25.63 $^{+0.10}_{-0.08}$	\\[4pt]
	NGC 300		&	26.34 $\pm$	0.06 &	e   &	26.60 $^{+0.06}_{-0.05}$	\\[4pt]
	NGC 55		&	26.84 $\pm$	0.07 &  f	&	26.62 $^{+0.03}_{-0.03}$	\\[4pt]
	NGC 2403	&	27.38 $\pm$	0.08 &  g	&	27.52 $^{+0.05}_{-0.05}$	\\[4pt]
	M81			&	27.73 $\pm$	0.04 &  h	&	27.84 $^{+0.09}_{-0.09}$	\\[4pt]
	M83			&	28.50 $\pm$	0.11 &  i	&	28.45 $^{+0.04}_{-0.03}$	\\[4pt]
	NGC 3621	&	28.96 $\pm$	0.14 &  j	&	29.22 $^{+0.09}_{-0.09}$	\\[4pt]
	M101		&	29.06 $\pm$	0.08 &  k	&	29.12 $^{+0.07}_{-0.08}$	\\[4pt]
	NGC 4258	&	29.49 $\pm$	0.07 &  l	&	29.42 $^{+0.03}_{-0.04}$	\\
	\\[-2.5ex]
	\enddata
	\tablecomments{The references to the original \fglr\ data are given below. The \fglr\ distances are homogenized to the		same calibration and zero point. The \trgb\ distances are extracted from the Extragalactic Distance Database (\citealt{Tully:2009, Anand:2021}).	}
	\tablerefs{$^a$\citet{Berger:2018}; $^b$\citet{Liu:2022}; $^c$\citet{Urbaneja:2008}; $^d$\citet{Hosek:2014};		$^e$\citet{Kudritzki:2008}; $^f$\citet{Kudritzki:2016}; $^g$\citet{Bresolin:2022}; $^h$\citet{Kudritzki:2012}; $^i$\citet{Bresolin:2016}; $^j$\citet{Kudritzki:2014}; $^k$This paper; $^l$\citet{Kudritzki:2024}.}	
\end{deluxetable}

\section{An independent determination of $H_0$} \label{sec:Hubble}

M101 is the host galaxy of the SN~Ia 2011fe. We can therefore
utilize the distance we obtained in Section~\ref{sec:distance} from the \fglr, calibrated via the geometric distance to the LMC, as the intermediate rung of the distance ladder which, together with the standardized magnitude of this single supernova event, allows us to derive the value of $H_0$.

According to \citet{Riess:2022}, the standardized $B$ magnitude of SN~2011fe is $m_B = 9.78\pm0.115$. This value is adopted from the Pantheon+ analysis by \citet{Brout:2022} and \citet{Scolnic:2022}.  With a \fglr\ distance modulus of $29.06\pm0.08$ mag, this leads to an absolute magnitude $M_B = -19.28\pm0.14$. Using the simple approach outlined in \citet[their Equation~6]{Riess:2024}, we have

\begin{equation}
    5\log\left({H_0 \over 72.5}\right) = M_B +19.28,
\end{equation}

\noindent
which yields $H_0 = 72.5\pm4.6$ km\,s$^{-1}$\,Mpc$^{-1}$.
For comparison, \citet{Riess:2022} obtained $H_0 = 73.04\pm1.04$ km\,s$^{-1}$\,Mpc$^{-1}$ from a Cepheid-based calibration of 42 SNe~Ia. Our determination is based on one supernova host only and, consequently, the error is relatively large. However, we point out that this result is an independent new approach, alternative to methods using Cepheids, \trgb, Mira stars, surface brightness fluctuations, or interstellar masers. Improving on our \fglr-based result will require acquiring \bsg\ spectra at considerably larger distances than M101, potentially with the next generation of 30m-class telescopes equipped with adaptive optics. Enlarging the sample to four galaxies will cut the error in half and provide crucial independent information on the Hubble tension.

\section{Summary} \label{sec:summary}
We have collected optical spectra of 42 blue supergiant (\bsg) candidates in the spiral galaxy M101, utilizing the LRIS instrument on the Keck~I telescope. We have carried out a quantitative spectral analysis for a subsample of 13 late-B and A type stars, determining surface gravities, effective temperatures, metallicities $Z$ and flux-weighted gravities.

The galactocentric radial metallicity gradient that we measure using the \bsg\ data is in agreement, within the uncertainties, with the gradient obtained from \hii\ regions previously studied in this galaxy with the direct method, once we account for a modest depletion of oxygen -- the chemical element representing the gas-phase metallicity -- on dust grains.
We also confirm our earlier finding that the \bsg\ metallicities agree well with the nebular abundances obtained with the \citet{Pettini:2004} O3N2 strong-line method.

We derive a spectroscopic distance to M101, relying on the calibrated relation between stellar parameters and intrinsic luminosity of \bsg s, the flux-weighted gravity--luminosity relationship (\fglr). The value that we obtain, $D=6.5\pm0.2$ Mpc ($\mu = 29.06 \pm 0.08$), is within the lower bound of recent ground- and space-based determinations from other techniques, but still lying within $\sim$1$\sigma$ from measurements based on both \trgb\ and Cepheids. 

Considering our new results on M101 within the broader context of \bsg\ studies in nearby galaxies, we find that the metallicities of star-forming galaxies, when calculated with the direct method and adjusted upwards for a $\sim$0.15 dex oxygen dust depletion factor, are in good agreement with those we infer from
the \bsg s, over a range in $Z$ of approximately 1.7 dex. 
Attributing the slight trend in the offset between stellar and nebular metallicities with $Z$ to the metallicity dependence of dust depletion, we derive an expression for the dust depletion factor of oxygen in \hii\ regions. We also update the galaxy mass-metallicity relation obtained from our project on young massive stars in external galaxies.

We show that the Cepheid metallicities, used by Riess and collaborators to calculate the distance to M101, are in substantial agreement with those of the blue supergiants. The small systematic offset produces a negligible effect on the distance.
Finally, exploiting the fact that M101 is the nearest SN~Ia host, we determine an independent value of the Hubble constant $H_0 = 72.5\pm4.6$ km\,s$^{-1}$\,Mpc$^{-1}$, which is based on our \fglr\ distance to this galaxy.

\begin{acknowledgments}
This research has made use of the Keck Observatory Archive (KOA), which is operated by the W. M. Keck Observatory and the NASA Exoplanet Science Institute (NExScI), under contract with the National Aeronautics and Space Administration.
This research has made use of the NASA/IPAC Extragalactic Database (NED),
which is operated by the Jet Propulsion Laboratory, California Institute of Technology,
under contract with the National Aeronautics and Space Administration.
RPK and ES acknowledge support by the Munich Excellence Cluster Origins and the Munich Institute for Astro-, Particle and Biophysics (MIAPbP) both funded by the Deutsche
Forschungsgemeinschaft (DFG, German Research Foundation) under Germany's Excellence Strategy EXC-2094 390783311. In addition, ES acknowledges support by the COMPLEX project from the European Research Council (ERC) under the European Union's Horizon 2020 research and innovation program grant agreement ERC-2019-AdG 882679.
The authors wish to recognize and acknowledge the very significant cultural role and reverence that the summit of Maunakea has always had within the indigenous Hawaiian community. We are most fortunate to have the opportunity to conduct observations from this mountain.
\end{acknowledgments}

\facility{Keck:I (LRIS)}

\software{IRAF (\citealt{Tody:1986, Tody:1993}), SciPy (\citealt{Virtanen:2020}), NumPy (\citealt{Harris:2020}), Matplotlib (\citealt{Hunter:2007a}), PyRAF (\citealt{Science-Software-Branch-at-STScI:2012}).}


\end{document}